\documentclass[superscriptaddress,twocolumn,showpacs,prb]{revtex4-2}
\usepackage[utf8]{inputenc}
\usepackage{amsmath}
\usepackage{mathtools}
\usepackage{braket}
\usepackage{graphicx}
\usepackage{amsfonts}
\usepackage{pgfplots}
\usepackage{csquotes}
\usepackage{hhline}
\usepackage{amssymb}
\usepackage{listings}
\usepackage{color}
\usepackage{xcolor}

\usepackage{tikz}
\usetikzlibrary{quantikz}

\makeatletter
\newcommand{\ssymbol}[1]{^{\@fnsymbol{#1}}}
\makeatother

\usepackage{lipsum}
\definecolor{codegreen}{rgb}{0,0.6,0}
\definecolor{codegray}{rgb}{0.5,0.5,0.5}
\definecolor{codepurple}{rgb}{0.58,0,0.82}
\definecolor{backcolour}{rgb}{0.95,0.95,0.92}
 
\lstdefinestyle{mystyle}{
    backgroundcolor=\color{backcolour},   
    commentstyle=\color{codegreen},
    keywordstyle=\color{magenta},
    numberstyle=\tiny\color{codegray},
    stringstyle=\color{codepurple},
    basicstyle=\footnotesize,
    breakatwhitespace=false,         
    breaklines=true,                 
    captionpos=b,                    
    keepspaces=true,                 
    numbers=left,                    
    numbersep=5pt,                  
    showspaces=false,                
    showstringspaces=false,
    showtabs=false,                  
    tabsize=2
}
 
\usepackage{subfigure}

\usepackage{dcolumn}
\usepackage{tabularx}
\usepackage[colorlinks=true,linkcolor=blue,citecolor=blue,urlcolor=blue]{hyperref}
\usepackage{longtable}
\usepackage{braket}
\usepackage{float}
\newcolumntype{C}{>{\centering\arraybackslash}X}
\usepackage{multirow}

\usepackage{tikz,xcolor,hyperref}
\definecolor{lime}{HTML}{A6CE39}
\DeclareRobustCommand{\orcidicon}{%
	\begin{tikzpicture}
	\draw[lime, fill=lime] (0,0) 
	circle [radius=0.16] 
	node[white] {{\fontfamily{qag}\selectfont \tiny ID}};
	\draw[white, fill=white] (-0.0625,0.095) 
	circle [radius=0.007];
	\end{tikzpicture}
	\hspace{-2mm}}
\foreach \x in {A, ..., Z}{%
	\expandafter\xdef\csname orcid\x\endcsname{\noexpand\href{https://orcid.org/\csname orcidauthor\x\endcsname}{\noexpand\orcidicon}}}

\pgfplotsset{compat=1.17}

\begin{document}
\title{Complexity analysis of quantum teleportation via different entangled channels in the presence of noise}

\author{Deepak Singh \orcidA{}}
\email{dsingh@ma.iitr.ac.in}
\affiliation{Department of Mathematics,\\ Indian Institute of Technology Roorkee 247667, Uttarakhand, India}

\author{Sanjeev Kumar \orcidB{}}
\email{sanjeev.kumar@ma.iitr.ac.in}
\affiliation{Department of Mathematics,\\ Indian Institute of Technology Roorkee 247667, Uttarakhand, India}

\author{Bikash K. Behera \orcidC{}}
\email{bikash@bikashsquantum.com}
\affiliation{Bikash's Quantum (OPC) Pvt. Ltd., Balindi, Mohanpur 741246, Nadia, West Bengal, India}
                                                                        
\begin{abstract}
Quantum communication is one of the hot topics in quantum computing, where teleportation of a quantum state has a slight edge and gained significant attention from researchers. A large number of teleportation schemes have already been introduced so far. Here, we compare the teleportation of a single qubit message among different entangled channels such as the two-qubit Bell channel, three-qubit GHZ channel, two- and three-qubit cluster states, the highly entangled five-qubit Brown \emph{et al.} state and the six-qubit Borras \emph{et al.} state. We calculate and compare the quantum costs in each of the cases. Furthermore, we study the effects of six noise models, namely bit-flip noise, phase-flip noise, bit-phase flip noise, amplitude damping, phase damping and the depolarizing error that may affect the communication channel used for the teleportation. An investigation on the variation of the initial state's fidelity with respect to the teleported state in the presence of the noise model is performed. A visual representation of the variation of fidelity for various values of the noise parameter $\eta$ is done through a graph plot. It is observed that as the value of noise parameter in the range $\eta \in [0,0.5]$, the fidelity decreases in all the entangled channels under all the noise models. After that, in the Bell channel, GHZ channel and three-qubit cluster state channel, the fidelity shows an upward trend under all the noise models. However, in the other three channels, the fidelity substantially decreases in the case of amplitude damping, phase damping and depolarizing noise, and even it reaches zero for $\eta = 1$ in Brown \emph{et al.} and Borras \emph{et al.} channels.
\end{abstract}


\maketitle

\section{Introduction}\label{sec:introduction}
Quantum communication deals with the transmission of quantum state between distance sites. It plays a prominent role in the quantum information processing \cite{DMW2007, NCCUP2000}. Quantum communication is getting improved day by day with the improvement in current technological advancement. Quantum networking has a broad spectrum of applications, quantum devices such as quantum routers \cite{YuanSR2015,ErhardQST2017,Bartkiewicz,BeheraQIP2019}, quantum repeaters \cite{BriegelEEPQS1999,MunroIEEE2015,AzumaNC2015,Behera2QIP2019,RuihongJPCS2019}, quantum satellite have been developed recently by China \cite{Vallone2015, Gibney2016, Wanisch2020}, which claims itself to be secure. Security is an essential aspect of today's communication; in the present scenario, most of the data is shared and stored on the cloud, so there is always a threat to break into the data; for which high security of the data is essential. Many quantum communication protocols have been developed so far such as quantum cryptography \cite{gisin2002, bennett1992, bennett1992quantum}, quantum information sharing \cite{das2013experimental,zheng2006splitting, muralidharan2008quantum}, quantum superdense coding \cite{harrow2004, xia2007controlled}, quantum remote state preparation \cite{bennett2005remote, n.b.2021}, quantum hierarchical remote state preparation \cite{shukla2017hierarchical, barik2020deterministic}, and quantum teleportation \cite{BennettPRL1993,Bouwmeester1997, Wang2015, Metcalf2014} to name a few.

Quantum teleportation is a process of transmitting quantum information from one place to another via a shared entangled channel, which in classical physics can be obtained by measurement. Entanglement plays a key role in the process of quantum teleportation because of its non-local properties. When an entangled quantum state is shared between two or more parties, they can teleport a message using the entangled channel. They have to apply certain unitary operations on appropriate qubits to teleport the message to a particular qubit. The first time quantum teleportation was proposed by Bennett \emph{et al.} \cite{BennettPRL1993} in 1993 via Einstein-Podolsky-Rosen (EPR). After then many different protocols are put forward with the extension of qubits \cite{RG2019}, dense coding in \cite{Hu2013}. Quantum teleportation is implemented experimentally in some of the research works presented \cite{Bouwmeester1997, Wang2015, Metcalf2014, Riebe2004} in various quantum systems, apart from that many theoretical and experimental quantum teleportation schemes have been proposed by researchers so far \cite{H2014, Nandi2014, Q2010, Nie2011, Muralidharan2008, Agrawal2006, Ursin2004}. Joo \emph{et al.} \cite{joo2003quantum} proposed two different schemes of teleporting a single qubit state by using W-state. Ghosh \emph{et al.} \cite{ghosh2002entanglement} have generalized a teleportation scheme of an entangle state using a Greenberger-Horne-Zeilinger(GHZ)-class state. Several new schemes are proposed for the teleportation using W-state and GHZ-state or GHZ-like state \cite{tsai2010teleportation}. The cluster states channel is used by Li and Cao \cite{da2007teleportation} to teleport arbitrary two states using a four-qubit cluster state. The teleportation of a three-qubit state using the five-qubit cluster sate is also performed by Liu \emph{et al.} \cite{liu2014quantum}. Then quantum teleportation of an arbitrary two-qubit state has been performed using a four-qubit cluster state \cite{RajiuddinQIP2020}. Quantum controlled teleportation of arbitrary qubit states have also been demonstrated via cluster states \cite{KumarSciRep2020}. Then a new way of quantum teleportation has been proposed using entanglement in coined quantum walk \cite{CarneiroNJP2005,WangQIP2017,ChatterjeeQIP2020}, which has taken tremendous attention of the scientific community. Sangchul \emph{et al.} \cite{oh2002fidelity} have studied the fidelity of quantum teleportation under the noisy channels. An experimental study of the teleportation under the effect of noise is done by Laura \emph{et al.} \cite{knoll2014noisy}.

In this study, we have made a complexity analysis of teleportation of a single qubit message using different entangled channels such as the Bell channel, three-qubits GHZ channel, two-qubit cluster state, three-qubit cluster state, highly entangled five-qubit Brown \emph{et al.} state and the six-qubit Borras \emph{et al.} state. Here, we apply quantum gates on the circuit to perform the teleportation. As the number of quantum gates increases in a protocol, it increases the complexity of teleportation. Therefore, we have calculated the quantum cost for all the above teleportation protocols and a graphical histogram representation is made in Fig. \ref{fig:Q Cost}  to compare the quantum cost of different teleportation protocols. Any real quantum experiment cannot be performed without the presence of noise in the channel. Noise is an inevitable feature of quantum communication that affects teleportation. The noise in the quantum system changes its state from a pure state to a mixed state, and therefore some quantum information is lost during the process of teleportation. Thus, we have introduced some noisy environments in each of the entangled channels used for teleportation. We have taken the six types of noises that may affect the teleportation process, namely the bit-flip noise, phase-flip noise, bit-phase flip noise, amplitude damping, phase damping and depolarizing noise. We have analytically derived the density matrices for all the entangled channels under every noise model. As fidelity gives the closeness between the two quantum states, we have calculated the fidelity between the initial and teleported states under noise. Then, a visual representation is made by plotting a graph between the noise parameter and the fidelity.

The paper is organized in following manner: Section \ref{Sec1.5} contains some basic definitions, sections \ref{Sec2}, \ref{Sec3}, \ref{Sec4}, \ref{Sec5}, \ref{Sec6}, and \ref{Sec7} discuss the teleportation process via the Bell channel, via GHZ channel, via the two-qubit cluster state channel, via three-qubit cluster state channel, via Brown \emph{et al.} channel, and via Borras \emph{et al.} channel respectively. Then in Section \ref{Sec8}, the analysis of noise in the entangled channel used for teleportation are performed. Finally, we conclude in Section \ref{Sec9} contains some discussion and conclusion followed by discussing the future works.

\section{Basic Definitions}\label{Sec1.5}
\subsection{Quantum Cost}
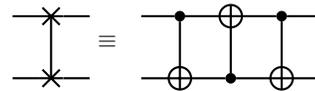
\begin{figure}[!htb]
\centering
\begin{quantikz}[column sep=0.35cm]
 \qw & \swap{1} & \qw  \\
 \qw & \targX{} & \qw 
\end{quantikz}
$\equiv$
\begin{quantikz}[column sep=0.35cm]
& \ctrl{1} & \targ{} & \ctrl{1} & \qw \\
& \targ{} & \ctrl{-1} & \targ{} & \qw
\end{quantikz}
    
\caption{SWAP gate decomposition}
\label{fig:SWAP}
\end{figure}

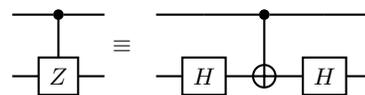
\begin{figure}[!htb]
\centering
\begin{quantikz}[column sep=0.35cm]
 \qw & \ctrl{1} & \qw  \\
 \qw & \gate{Z} & \qw 
\end{quantikz}
$\equiv$
\begin{quantikz}[column sep=0.35cm]
& \qw & \ctrl{1} & \qw  & \qw \\
& \gate{H} & \targ{} & \gate{H} & \qw
\end{quantikz}
    
\caption{Control-Z gate decomposition}
\label{fig:CZ}
\end{figure}

Quantum cost is the number of primitive reversible gates used in designing a circuit. The quantum cost of all basic single-qubit gates and control not is taken as unity \cite{soeken2010window}. The important property of every reversible gate is that it can be generated from a combination of basic quantum gates, so the quantum cost of every reversible gate can be calculated by counting the number of basic gates used in it. The Hadamard and control-not(CNOT) gate have a quantum cost equal to one. The other gate used in this paper is the control-Z gate and SWAP gates. These gates can be prepared with the help of Hadamard and CNOT gates given in Fig. \ref{fig:SWAP} and \ref{fig:CZ} respectively. The more is the quantum cost, and the more complexity will be in the execution of the circuit on the real quantum hardware. So in the next section, we have evaluated the quantum cost of each teleportation protocol via different entangled channels.

\section{Teleportation Schemes}
\subsection{Teleportation using Bell Channel}\label{Sec2}
An arbitrary single-qubit quantum message can be written as $\ket{M} = \alpha \ket{0} +\beta \ket{1}$, where $\alpha,\beta \in \mathbb{C}$ and $|\alpha|^2 + |\beta|^2 = 1$. A more convenient representation of the single qubit state is given by $\ket{M} = cos(\theta/2) \ket{0} + e^{i\phi}sin(\theta/2) \ket{1}$, where $0\leq \theta \leq \pi$ and $0\leq \phi < 2\pi$. Now, the teleportation of this single qubit message is carried out using the two-qubit Bell channel given by $\ket{\psi^{\pm}} = \frac{1}{\sqrt{2}} (\ket{00} \pm \ket{11} )$, $\ket{\phi^{\pm}} = \frac{1}{\sqrt{2}} (\ket{01} \pm \ket{10})$. Here, the entangled Bell channel is shared between two parties say Alice and Bob. Alice has the first qubit and Bob has the second qubit that form the entangled Bell channel. Now, let us take one of the Bell channels for teleportation process say $\ket{\psi^+}$, the combined state after taking the tensor product of $\ket{M}$ and $\ket{\psi^+}$ is given by Eq. \eqref{eq:bell},

\begin{eqnarray}
    && \ket{\xi} = \ket{M}_1 \otimes \ket{\psi^{+}}_{23} \nonumber\\ 
    && ~~~~~ = (\alpha \ket{0} + \beta \ket{1})_1 \otimes \Big( \frac{1}{\sqrt{2}}(\ket{00} + \ket{11} )_{23} \Big)\nonumber\\
    && ~~~~~ =  \frac{1}{\sqrt{2}} \Big( \alpha \ket{000} + \beta \ket{100} + \alpha \ket{011} + \beta \ket{111} \Big)_{123}
\label{eq:bell}    
\end{eqnarray}

Now, Alice applies the Bell basis measurement on the first and second qubit, i.e. $CX(1,2)$ and $H(1)$, the final state is shown in Eq. \eqref{eq:bell final},

\begin{eqnarray}
    && \ket{\xi} = \frac{\ket{00}_{12}}{2} \Big(\alpha \ket{0} + \beta \ket{1}\Big)_3 + 
    \frac{\ket{01}_{12}}{2} \Big(\alpha \ket{1} + \beta \ket{0}\Big)_3 \nonumber \\
    && ~~~~~ + \frac{\ket{10}_{12}}{2} \Big(\alpha \ket{0} - \beta \ket{1}\Big)_3 + 
    \frac{\ket{11}_{12}}{2} \Big(\alpha \ket{1} - \beta \ket{0}\Big)_3 \nonumber \\
\label{eq:bell final}    
\end{eqnarray}

\begin{figure}[!htb]
    \centering
    \includegraphics[width=0.5\textwidth]{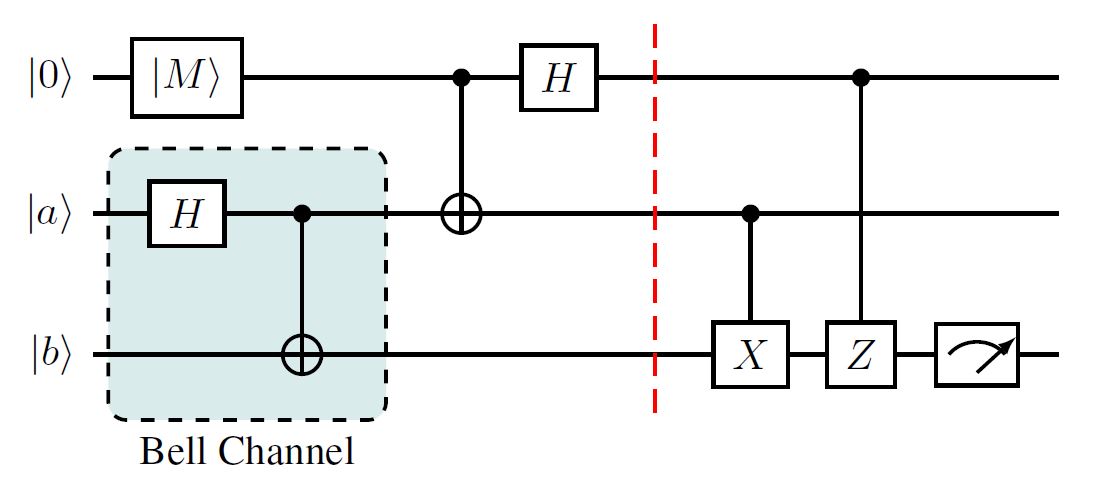}
    \caption{Quantum circuit for teleporting the message $\ket{M} = \alpha \ket{0} +\beta \ket{1}$ through the Bell channels, controlled operations are represented for $\ket{\psi_1}$ (given in table \ref{table:Bell}) after the red-dashed line, which will vary for different Bell-channels.}
    \label{fig:Bell_Circuit}
\end{figure}

The teleportation of the single qubit message $\ket{M}=\alpha \ket{0}+\beta \ket{1}$ at this stage has been done successfully to the Bob's qubit. However, now Bob needs to apply appropriate unitary operations on his qubit to recover the teleported message. The teleportation protocol is prepared in the following quantum circuit given in Fig. \ref{fig:Bell_Circuit}. The controlled operations are represented in the circuit \ref{fig:Bell_Circuit} after the red dash line. Here in the quantum circuit, $\ket{a},\ket{b} \in \{ \ket{0},\ket{1}\}$. Choosing different values of $\ket{a}$ and $\ket{b}$ we get different Bell channels shown in Table \ref{table:Bell}.

\begin{table}[!htb]
\centering
\renewcommand{\arraystretch}{1.65}
\begin{tabular}{|l|l|l|c|}
\hline
\textbf{$\ket{a}$} & \textbf{$\ket{a}$} & \multicolumn{1}{c|}{\textbf{Bell States}} & \textbf{Quantum Cost} \\ \hline
$\ket{0}$ & $\ket{0}$ & $\ket{\psi^{+}} = \frac{1}{\sqrt{2}} (\Ket{00} + \Ket{11} )$ & 9 \\ \hline
$\ket{0}$ & $\ket{1}$ & $\ket{\phi^{+}} = \frac{1}{\sqrt{2}} (\Ket{01} + \Ket{10} )$ & 11 \\ \hline
$\ket{1}$ & $\ket{0}$ & $\ket{\psi^{-}} = \frac{1}{\sqrt{2}} (\Ket{00} - \Ket{11} )$ & 11 \\ \hline
$\ket{1}$ & $\ket{1}$ & $\ket{\phi^{-}} = \frac{1}{\sqrt{2}} (\Ket{01} - \Ket{10} )$ & 13 \\ \hline
\end{tabular}
\caption{All possible Bell channels for choosing different values of $\ket{a}$ and $\ket{b}$. The last column contains the quantum cost for teleporting single qubit message via different Bell channels.}
\label{table:Bell}
\end{table}

The Bell basis measurement on the first two qubits is done by applying a CNOT gate from the first to the second qubit, i.e. $CX(1,2)$, followed by a Hadamard gate on the first qubit, i.e. $H(1)$. After Alice has performed the Bell basis measurement on the first two qubits, she communicates her measurement result to Bob through a classical channel. Now, Bob applies the appropriate unitary operation on his qubit to recover the message and complete the teleportation process. To equivalently achieve the above same scenario, Bob applies control operations from Alice's qubits to his qubit. The controlled operations in case of Bell channel $\ket{\psi_+}=\frac{1}{\sqrt{2}}(|00\rangle+|11\rangle)$ are given in the quantum circuit (Fig. \ref{fig:Bell_Circuit}) after the red dashed line. For all the cases, the controlled unitary operations are shown in Table \ref{table:Bell Operations}.

\begin{table}[!htb]
\centering
\setlength{\tabcolsep}{4.5pt}
\renewcommand{\arraystretch}{1.5}
\begin{tabular}{|l|l|l|l|l|}
\hline
\multirow{2}{*}{\textbf{Alice's Measurement}} & \multicolumn{4}{c|}{\textbf{Control operations}} \\ \cline{2-5} 
& $\ket{\psi^{+}}$  & $\ket{\psi^{-}}$ & $\ket{\phi^{+}}$ & $\ket{\phi^{-}}$ \\ \hline
$\ket{00}$ & $\mathbb{I}$ & $\mathbb{Z}$ & $\mathbb{X}$ & $\mathbb{ZX}$ \\ \hline
$\ket{01}$ & $\mathbb{X}$ & $\mathbb{ZX}$ & $\mathbb{I}$ & $\mathbb{Z}$  \\ \hline
$\ket{10}$ & $\mathbb{Z}$ & $\mathbb{I}$  & $\mathbb{ZX}$ & $\mathbb{X}$ \\ \hline
$\ket{11}$ & $\mathbb{ZX}$ & $\mathbb{X}$ & $\mathbb{Z}$ & $\mathbb{I}$ \\ \hline
\end{tabular}
\caption{Controlled operations that Bob needs to apply to recover the teleported message $\ket{M} = \alpha \ket{0} +\beta \ket{1}$ via all four Bell channels.\label{table:Bell Operations}}
\end{table}


\subsection{Teleportation using GHZ Channel}\label{Sec3}
Here, the underlying entangled channel is the three-qubit GHZ channel. The teleportation is performed through this channel, which is given by $\ket{\psi_i} = \frac{1}{\sqrt{2}} (\ket{x_1 x_2 x_3} \pm \ket{\Bar{x_1}\Bar{x_2}\Bar{x_3}} )_{123}$, where $x_1,x_2,x_3 \in \{0,1\}$ and $\Bar{x_1},\Bar{x_2},\Bar{x_3}$ are their respective conjugates. Here, the first qubit belongs to Alice, second qubit belongs to another controller Charlie and third qubit belongs to Bob. Now, to teleport the secret message $\ket{M} = \alpha \ket{0} +\beta \ket{1}$ through one of the GHZ channels say, $\ket{\psi_1} = \frac{1}{\sqrt{2}} (\ket{000} + \ket{111} )_{123}$, the combined state after taking the tensor product of $\ket{M}$ and $\ket{\psi_1}$ is given by Eq. \eqref{eq:GHZ},
\begin{eqnarray}
    \ket{\xi} &=& \ket{M}_1 \otimes \ket{\psi_{1}}_{234} \nonumber \\ 
              &=& (\alpha \ket{0} + \beta \ket{1})_1 \otimes \left(\frac{1}{\sqrt{2}} (\ket{000} + \ket{111} )_{234} \right)
\label{eq:GHZ}    
\end{eqnarray}

Now, Alice applies the Bell basis measurement on the first and second qubit, i.e. $CX(1,2)$ and $H(1)$, Charlie applies Hadamard gate on the third qubit $H(3)$, the final state is shown in Eq. \eqref{eq:GHZ final}.

\begin{eqnarray}
&& \ket{\xi} = 
    \frac{\ket{00}_{12}}{2} \Big( \ket{0}_3 (\alpha \ket{0} + \beta \ket{1})_4 + \ket{1}_3 (\alpha \ket{0} - \beta \ket{1})_4 \Big)  \nonumber \\
    &&+ 
    \frac{\ket{01}_{12}}{4} \Big( \ket{0}_3 (\alpha \ket{1} + \beta \ket{0})_4 + \ket{1}_3 (-\alpha \ket{1} + \beta \ket{0})_4 \Big)  \nonumber\\
    &&+ 
    \frac{\ket{10}_{12}}{4}\Big( \ket{0}_3 (\alpha \ket{0} - \beta \ket{1})_4 + \ket{1}_3 (\alpha \ket{0} + \beta \ket{1})_4 \Big) \nonumber \\ 
    &&+
    \frac{\ket{11}_{12}}{4}\Big( \ket{0}_3 (\alpha \ket{1} - \beta \ket{0})_4 + \ket{1}_3 (-\alpha \ket{1} - \beta \ket{0})_4 \Big)\nonumber\\
\label{eq:GHZ final}    
\end{eqnarray}

\begin{figure}[!htb]
    \centering
    \includegraphics[width=0.5\textwidth]{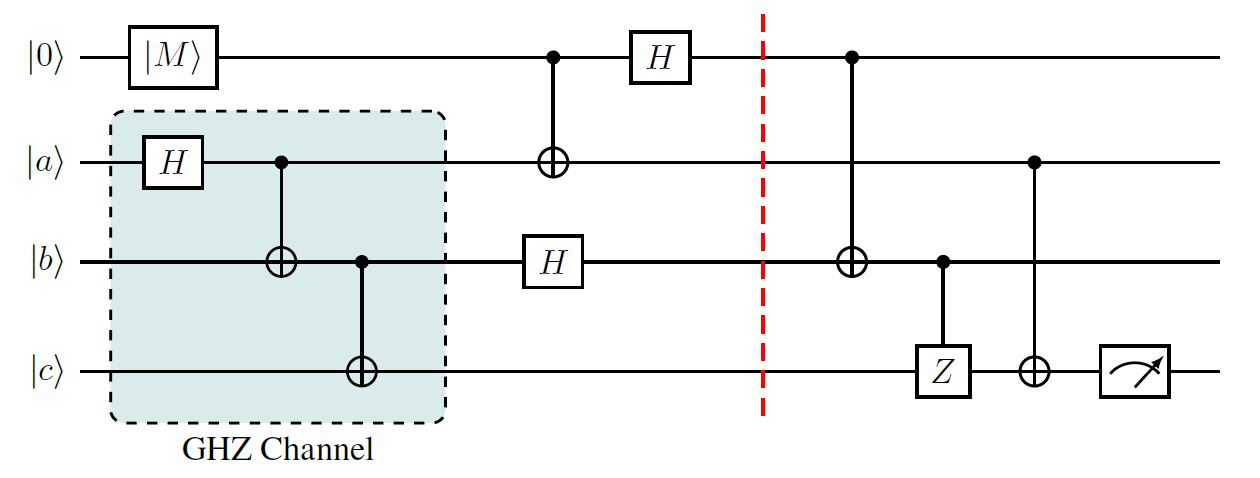}
    \caption{Teleportation protocol of a single qubit message $\ket{M} = \alpha \ket{0} +\beta \ket{1}$ using entangled GHZ channel, controlled operations are represented for $\ket{\psi_1}$ (given in table \ref{table:GHZ States}) after the red-dashed line, which will vary for different GHZ-channels.}
    \label{fig:GHZ Channel}
\end{figure}

Here, the single-qubit message $\ket{M}=\alpha \ket{0} + \beta \ket{1}$ has been teleported to Bob's qubit successfully. Now, before making a measurement, Bob needs to apply proper unitary operations on his qubit to recover the message. The unitary operations are shown in Table \ref{table:GHZ Operation}. The teleportation protocol via GHZ channel is demonstrated via the following quantum circuit (Fig. \ref{fig:GHZ Channel}), where $\ket{a},\ket{b}, \ket{c}\in \{ \ket{0},\ket{1}\}$, choosing different combinations of values of $\ket{a},\ket{b}$ and $\ket{c}$, all of the eight three-qubit GHZ states are shown in Table \ref{table:GHZ States}. To recover the teleported messages, Bob needs to apply certain controlled unitary operations on his qubit. For all of the eight entangled GHZ channels, the controlled operations are given in Table \ref{table:GHZ Operation}. \\

\begin{table}
\centering
\renewcommand{\arraystretch}{1.75}
\begin{tabular}{|l|l|l|l|c|}
\hline
\textbf{$\ket{a}$} & \textbf{$\ket{a}$} & \textbf{$\ket{a}$} & \multicolumn{1}{c|}{\textbf{GHZ State}} & \textbf{Quantum Cost} \\ \hline
$\ket{0}$ & $\ket{0}$ & $\ket{0}$ & $\Ket{\psi_1}=\frac{1}{\sqrt{2}}(\Ket{000}+\Ket{111})$ & 12 \\ \hline
$\ket{0}$ & $\ket{0}$ & $\ket{1}$ & $\Ket{\psi_2}=\frac{1}{\sqrt{2}}(\Ket{001}+\Ket{110})$ & 13 \\ \hline
$\ket{0}$ & $\ket{1}$ & $\ket{0}$ & $\Ket{\psi_3}=\frac{1}{\sqrt{2}}(\Ket{010}+\Ket{101})$ & 13 \\ \hline
$\ket{0}$ & $\ket{1}$ & $\ket{1}$ & $\Ket{\psi_4}=\frac{1}{\sqrt{2}}(\Ket{000}-\Ket{111})$ & 14 \\ \hline
$\ket{1}$ & $\ket{0}$ & $\ket{0}$ & $\Ket{\psi_5}=\frac{1}{\sqrt{2}}(\Ket{011}+\Ket{100})$ & 15 \\ \hline
$\ket{1}$ & $\ket{0}$ & $\ket{1}$ & $\Ket{\psi_6}=\frac{1}{\sqrt{2}}(\Ket{001}-\Ket{110})$ & 18 \\ \hline
$\ket{1}$ & $\ket{1}$ & $\ket{0}$ & $\Ket{\psi_7}=\frac{1}{\sqrt{2}}(\Ket{010}-\Ket{101})$ & 15 \\ \hline
$\ket{1}$ & $\ket{1}$ & $\ket{1}$ & $\Ket{\psi_8}=\frac{1}{\sqrt{2}}(\Ket{011}-\Ket{100})$ & 20 \\ \hline
\end{tabular}
\caption{Three qubit GHZ states after choosing different combinations of $\ket{a}$, $\ket{b}$ and $\ket{c}$. The last column contains the quantum cost for teleporting a single qubit message via the GHZ states.}
\label{table:GHZ States}
\end{table}

\begin{table*}
\centering
\begin{tabular}{|l|l|l|l|l|l|l|l|l|l|}
\hline
\multicolumn{1}{|c|}{\multirow{2}{*}{\textbf{\begin{tabular}[c]{@{}c@{}}Alice's Measurement \\ (Qubit \{1,2\})\end{tabular}}}} &
  \multicolumn{1}{c|}{\multirow{2}{*}{\textbf{\begin{tabular}[c]{@{}c@{}}Charlie Measurement \\ (Qubit \{3\})\end{tabular}}}} &
  \multicolumn{8}{c|}{\textbf{Bob's Operation (Qubit\{4\})}} \\ \cline{3-10} 
\multicolumn{1}{|c|}{} &
  \multicolumn{1}{c|}{} &
  $\ket{\psi_1}$ &
  $\ket{\psi_2}$ &
  $\ket{\psi_3}$ &
  $\ket{\psi_4}$ &
  $\ket{\psi_5}$ &
  $\ket{\psi_6}$ &
  $\ket{\psi_7}$ &
  $\ket{\psi_8}$ \\ \hline
\multirow{2}{*}{$\ket{00}$} &
  $\ket{0}$ &
  $\mathbb{I}$ &
  $\mathbb{Z}$ &
  $\mathbb{X}$ &
  $\mathbb{ZX}$ &
  $\mathbb{I}$ &
  $\mathbb{Z}$ &
  $\mathbb{X}$ &
  $\mathbb{ZX}$ \\ \cline{2-10} 
 &
  $\ket{1}$ &
  $\mathbb{Z}$ &
  $\mathbb{I}$ &
  $\mathbb{ZX}$ &
  $\mathbb{X}$ &
  $\mathbb{Z}$ &
  $\mathbb{I}$ &
  $\mathbb{ZX}$ &
  $\mathbb{X}$ \\ \hline
\multirow{2}{*}{$\ket{01}$} &
  $\ket{0}$ &
  $\mathbb{X}$ &
  $\mathbb{ZX}$ &
  $\mathbb{I}$ &
  $\mathbb{Z}$ &
  $\mathbb{X}$ &
  $\mathbb{ZX}$ &
  $\mathbb{I}$ &
  $\mathbb{Z}$ \\ \cline{2-10} 
 &
  $\ket{1}$ &
  $\mathbb{ZX}$ &
  $\mathbb{X}$ &
  $\mathbb{Z}$ &
  $\mathbb{I}$ &
  $\mathbb{ZX}$ &
  $\mathbb{X}$ &
  $\mathbb{Z}$ &
  $\mathbb{I}$ \\ \hline
\multirow{2}{*}{$\ket{10}$} &
  $\ket{0}$ &
  $\mathbb{Z}$ &
  $\mathbb{I}$ &
  $\mathbb{ZX}$ &
  $\mathbb{X}$ &
  $\mathbb{Z}$ &
  $\mathbb{I}$ &
  $\mathbb{ZX}$ &
  $\mathbb{ZX}$ \\ \cline{2-10} 
 &
  $\ket{1}$ &
  $\mathbb{I}$ &
  $\mathbb{Z}$ &
  $\mathbb{X}$ &
  $\mathbb{ZX}$ &
  $\mathbb{I}$ &
  $\mathbb{Z}$ &
  $\mathbb{X}$ &
  $\mathbb{X}$ \\ \hline
\multirow{2}{*}{$\ket{11}$} &
  $\ket{0}$ &
  $\mathbb{ZX}$ &
  $\mathbb{X}$ &
  $\mathbb{Z}$ &
  $\mathbb{Z}$ &
  $\mathbb{ZX}$ &
  $\mathbb{X}$ &
  $\mathbb{Z}$ &
  $\mathbb{I}$ \\ \cline{2-10} 
 &
  $\ket{1}$ &
  $\mathbb{X}$ &
  $\mathbb{ZX}$ &
  $\mathbb{I}$ &
  $\mathbb{I}$ &
  $\mathbb{X}$ &
  $\mathbb{ZX}$ &
  $\mathbb{I}$ &
  $\mathbb{Z}$ \\ \hline
\end{tabular}
\caption{Controlled operations that Bob needs to apply to recover the teleported message $\ket{M}= \alpha \ket{0} + \beta \ket{1}$ via all the eight possible GHZ-channels. \label{table:GHZ Operation}}
\end{table*}

\subsection{Teleportation using two qubit Cluster state channel} \label{Sec4}
\begin{figure}[!htb]
    \centering
    \includegraphics[width=0.47\textwidth]{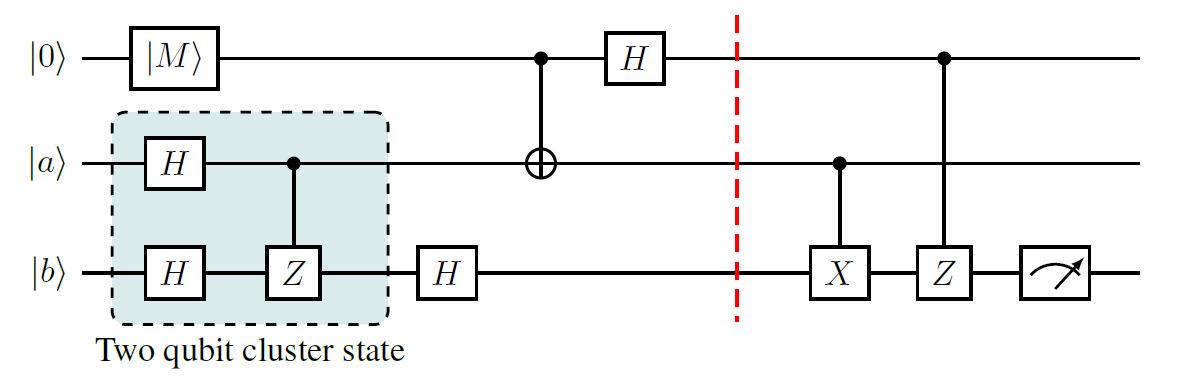}
    \caption{The teleportation protocol for teleporting a single qubit message $\ket{M}= \alpha \ket{0} + \beta \ket{1}$ via the two-qubit cluster state, controlled operations are represented for $\ket{\psi_1}$ (given in table \ref{table:2qClusterStateCost}) after the red-dashed line, which will vary for different two-qubit cluster states.}
    \label{fig:2q Channel tel}
\end{figure}

The two-qubit cluster state is an entangled channel shared by the parties Alice and Bob to perform the teleportation. It is given by $\ket{\psi_i}=(\ket{a}\ket{b})_{CZ(1,2)}$, where $\ket{a},\ket{b} \in \{\ket{+},\ket{-} \}$. The two-qubit cluster state channel is prepared inside the dashed-box in the circuit given in Fig. \ref{fig:2q Channel tel}. Taking $\ket{a}=\ket{+}$ and $\ket{b}=\ket{+}$, the two qubit cluster state is given by, $\ket{\psi_1}=(\ket{+}\ket{+})_{CZ(1,2)} = \frac{1}{2} \left( \ket{00} + \ket{01} + \ket{10} - \ket{11} \right)$.
Now, let us perform the teleportation protocol for the single qubit message $\ket{M} = \alpha \ket{0} +\beta \ket{1}$ via the two qubit cluster state, the combined state after taking the tensor product of $\ket{M}$ and $\ket{\psi_1}$ is given by Eq. \eqref{eq:2q Cluster}\\

\begin{eqnarray}
&& \ket{\xi} = \ket{M}_1 \otimes \ket{\psi_{1}}_{23} \nonumber\\ 
&&~~~~= (\alpha \ket{0} + \beta \ket{1})_1 \otimes \left( \frac{1}{2} ( \ket{00} + \ket{01} + \ket{10} - \ket{11}) \right)_{23} \nonumber\\ 
\label{eq:2q Cluster}    
\end{eqnarray}

Now, Alice applies the Bell basis measurement on her first and second qubit, i.e., $CX(1,2)$ , $H(1)$  and other operation $H(3)$ on the third qubit, given in circuit (Fig. \ref{fig:2q Channel tel}), the final state is shown in Eq. \eqref{eq:2q cluster_final},
 
\begin{eqnarray}
&&  \ket{\xi} =
    \frac{\ket{00}_{12}}{2} \left( \alpha \ket{0} + \beta \ket{1} \right)_3 + 
    \frac{\ket{01}_{12}}{2} \left( \alpha \ket{1} + \beta \ket{0} \right)_3 \nonumber\\
&& +\frac{\ket{10}_{12}}{2} \left( \alpha \ket{0} - \beta \ket{1} \right)_3 + 
    \frac{\ket{11}_{12}}{2} \left( \alpha \ket{1} - \beta \ket{0} \right)_3 \nonumber\\
\label{eq:2q cluster_final}    
\end{eqnarray}

At this stage of the protocol, teleportation of the message $\ket{M}=\alpha \ket{0} + \beta \ket{1}$ to Bob's qubit has been done successfully. Now, Bob has to apply appropriate unitary operations on his qubit to recover the message. The controlled operations that Bob need to apply on his qubit in case of $\ket{\psi_1} = \frac{1}{2} \left( \ket{00} + \ket{01} + \ket{10} - \ket{11} \right)$ are shown in quantum circuit (Fig. \ref{fig:2q Channel tel}) after the red colour dashed line. The unitary operations in all the other cases are shown in Table \ref{table:2q Cluster Op}. Here $\ket{a},\ket{b} \in \{ \ket{0},\ket{1}\}$. Choosing all combinations of $\ket{a}$ and $\ket{b}$, we get four types of two-qubit cluster state channels, shown in Table \ref{table:2qClusterStateCost}. Also, the teleportation is performed for each of the two-qubit cluster state channel and quantum cost is calculated, which is represented in the last column of Table \ref{table:2qClusterStateCost}. The controlled operations that Bob needs to apply on his qubit to recover the teleported message in all the two-qubit cluster state cases are shown in Table \ref{table:2q Cluster Op}.

\begin{table}[!htb]
\centering
\renewcommand{\arraystretch}{1.75}
\begin{tabular}{|l|l|l|c|}
\hline
\textbf{$\ket{a}$} & \textbf{$\ket{b}$} & \multicolumn{1}{c|}{\textbf{Two qubit cluster state}} & \textbf{Quantum Cost} \\ \hline
$\ket{0}$ & $\ket{0}$ & $\ket{\psi_1} = \frac{1}{2} (\ket{00} +\ket{01}+\ket{10}-\ket{11}$  & 13 \\ \hline
$\ket{0}$ & $\ket{1}$ & $\ket{\psi_2} = \frac{1}{2} (\ket{00} -\ket{01}+\ket{10}+ \ket{11}$ & 15 \\ \hline
$\ket{1}$ & $\ket{0}$ & $\ket{\psi_3} = \frac{1}{2} (\ket{00} +\ket{01}-\ket{10}+ \ket{11}$ & 15 \\ \hline
$\ket{1}$ & $\ket{1}$ & $\ket{\psi_4} = \frac{1}{2} (\ket{00} -\ket{01}-\ket{10}- \ket{11}$ & 17 \\ \hline
\end{tabular}
\caption{The two-qubit cluster states, after choosing different values of $\ket{a}$ and $\ket{b}$. The last column contains the quantum cost for teleporting a single qubit message via the two-qubit cluster states.}
\label{table:2qClusterStateCost}
\end{table}

\begin{table}[!htb]
\centering
\begin{tabular}{|l|l|l|l|l|}
\hline
\multirow{2}{*}{\textbf{Alice's Measurement}} & \multicolumn{4}{l|}{\textbf{Controlled Operations}} \\ \cline{2-5} 
           & $\ket{\psi_1}$ & $\ket{\psi_2}$ & $\ket{\psi_3}$ & $\ket{\psi_4}$ \\ \hline
$\ket{00}$ & $\mathbb{I}$   & $\mathbb{X}$   & $\mathbb{Z}$   & $\mathbb{Z}$   \\ \hline
$\ket{01}$ & $\mathbb{X}$   & $\mathbb{I}$   & $\mathbb{ZX}$  & $\mathbb{ZX}$  \\ \hline
$\ket{10}$ & $\mathbb{Z}$   & $\mathbb{ZX}$  & $\mathbb{I}$   & $\mathbb{X}$   \\ \hline
$\ket{11}$ & $\mathbb{ZX}$  & $\mathbb{Z}$   & $\mathbb{X}$   & $\mathbb{I}$   \\ \hline
\end{tabular}
\caption{Controlled operations that Bob needs to apply to recover the teleported message via the two-qubit cluster states.}
\label{table:2q Cluster Op}
\end{table}


\begin{figure}[!htb]
    \centering
    \includegraphics[width=0.45\textwidth]{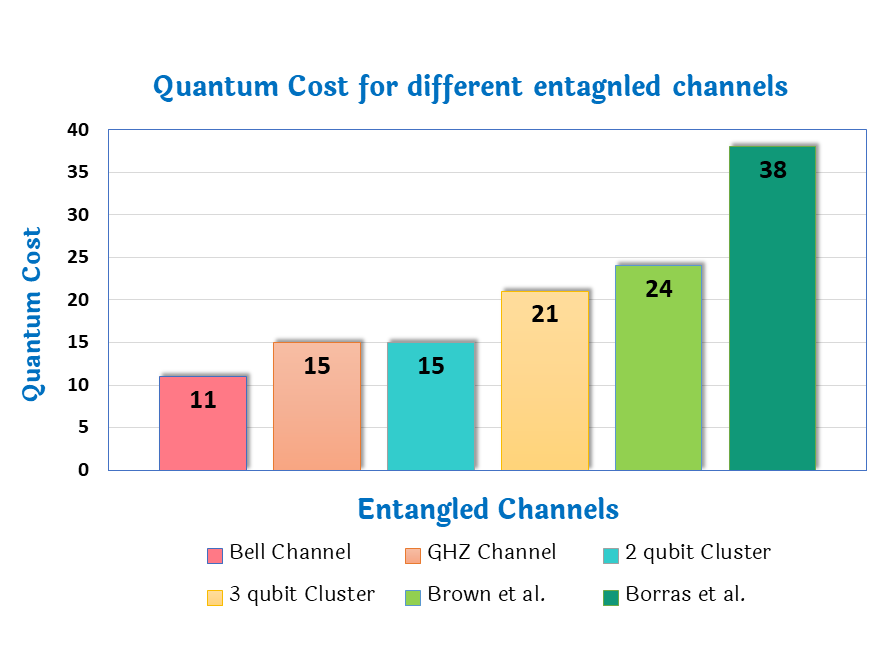}
    \caption{The quantum cost for different entangled channels. Average cost has been taken in case of Bell channel, GHZ channel, two-qubit cluster state and three-qubit cluster state channels.}
    \label{fig:Q Cost}
\end{figure}

\subsection{Teleportation using three-qubit cluster state channel}\label{Sec5}

The three-qubit cluster state shared by three parties Alice, Charlie and Bob. Here, Charlie acts as an extra controller in the teleportation. The three-qubit cluster state is given by $\ket{\psi}=(\ket{a}\ket{b}\ket{c})_{CZ(1,2)CZ(2,3)}$, where $\ket{a},\ket{b},\ket{c} \in \{\ket{+},\ket{-} \}$. Taking $\ket{a}=\ket{+}$, $\ket{b}=\ket{+}$ and $\ket{c}=\ket{+}$, one of the eight three-qubit cluster states is given in Eq. \eqref{eq:3qCeq},
\begin{eqnarray}
&& \ket{\psi_1} = (\ket{+}\ket{+}\ket{+})_{CZ(1,2)CZ(2,3)} \nonumber\\
&&~~~~~= \frac{1}{2\sqrt{2}} ( \ket{000} + \ket{001} 
+ \ket{010} - \ket{011} \nonumber \\ 
&&~~~~~+ \ket{100} + \ket{101} - \ket{110} + \ket{111} ) 
\label{eq:3qCeq} 
\end{eqnarray}

Now, to perform the teleportation protocol for the single-qubit message $\ket{\phi} = \alpha \ket{0} + \beta \ket{1}$ via the three-qubit cluster state channel, the combined state after taking the tensor product of $\ket{M}$ and $\ket{\psi_1}$ is given by Eq. \eqref{eq:3q Cluster} \\

\begin{eqnarray}
    \ket{\xi} &=& \ket{M}_1 \otimes \ket{\psi_{1}}_{234} \nonumber\\ 
              &=& (\alpha \ket{0} + \beta \ket{1})_1 \otimes  \frac{1}{2\sqrt{2}} ( \ket{000} + \ket{001} + \ket{010} \nonumber\\
              &-& \ket{011} + \ket{100} + \ket{101} - \ket{110} + \ket{111} )_{234}
\label{eq:3q Cluster}    
\end{eqnarray}

\begin{figure}[!htb]
    \centering
    \includegraphics[width=0.5\textwidth]{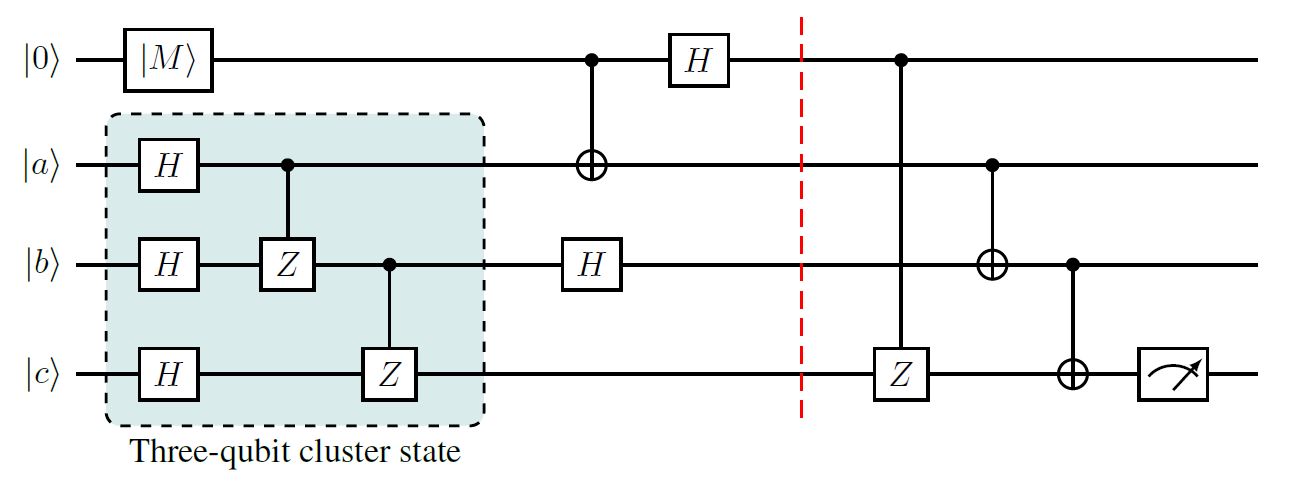}
    \caption{The teleportation protocol for teleporting a single qubit message $\ket{M}= \alpha \ket{0} + \beta \ket{1}$ via the three-qubit cluster state, controlled operations are represented for $\ket{\psi_1}$ (given in table \ref{table:3q cluster States}) after the red-dashed line, which will vary for different three-qubit cluster states.}
    \label{fig:3Qubit Cluster Channel}
\end{figure}

Now, Alice applies the Bell basis measurement on the first and second qubit, i.e. $CX(1,2)$, $H(1)$ and other operation $H(3)$ on the third qubit as depicted in quantum circuit given in Fig. \ref{fig:3Qubit Cluster Channel}, the final state is shown in Eq. \eqref{eq:3q cluster_final},

\begin{eqnarray}
&&  \ket{\xi} = 
    \frac{\ket{00}_{12}}{2} \Big( \ket{0}_3 (\alpha \ket{0} + \beta \ket{1})_4 + \ket{1}_3 (\alpha \ket{1} + \beta \ket{0})_4 \Big) \nonumber\\  
&&  +
    \frac{\ket{01}_{12}}{4} \Big( \ket{0}_3 (\alpha \ket{1} + \beta \ket{0})_4 + \ket{1}_3 (\alpha \ket{0} + \beta \ket{1})_4 \Big)\nonumber \\ 
&&  + 
    \frac{\ket{10}_{12}}{4} \Big( \ket{0}_3 (\alpha \ket{0} - \beta \ket{1})_4 + \ket{1}_3 (\alpha \ket{1} - \beta \ket{0})_4 \Big) \nonumber\\  
&&  +
    \frac{\ket{11}_{12}}{4} \Big( \ket{0}_3 (\alpha \ket{1} - \beta \ket{0})_4 + \ket{1}_3 (\alpha \ket{0} - \beta \ket{1})_4 \Big)\nonumber\\
\label{eq:3q cluster_final}    
\end{eqnarray}
 
After reaching this state, the message $\ket{M}=\alpha \ket{0} + \beta \ket{1}$ is successfully teleported to Bob's qubit. Now, Bob needs to apply the appropriate unitary operations on his qubit to recover the message. The unitary operations are shown in Table \ref{table:3q Cluster Operation}. The teleportation protocol is shown in the following quantum circuit (Fig. \ref{fig:3Qubit Cluster Channel}). Here $\ket{a},\ket{b}, \ket{c}\in \{ \ket{0},\ket{1}\}$. Choosing all the eight combinations of qubits $\ket{a}, \ket{b}$ and $\ket{c}$ we get eight different forms of three-qubit cluster states, which are shown in Table \ref{table:3q cluster States}. Also, the teleportation protocol is performed for each of the eight channels, and the quantum cost is calculated, which is represented in the last column of Table \ref{table:3q cluster States}. The controlled operations that Bob needs to apply on his qubit to recover the message in all the eight three-qubit cluster states entangled channels are given in Table \ref{table:3q Cluster Operation}. \\

\begin{table*}
\centering
\renewcommand{\arraystretch}{1.75}
\begin{tabular}{|l|l|l|l|c|}
\hline
\textbf{$\ket{a}$} & \textbf{$\ket{b}$} & \textbf{$\ket{c}$} & \multicolumn{1}{c|}{\textbf{Three Qubit Cluster States}} & \textbf{Quantum Cost} \\ \hline
$\ket{0}$ & $\ket{0}$ & $\ket{0}$ & $\Ket{\psi_1}=\frac{1}{\sqrt{2}}(\Ket{000}+\Ket{001}+\Ket{010}-\Ket{011}+\Ket{100}+\Ket{101}-\Ket{110}+\Ket{111})$ & 18 \\ \hline
$\ket{0}$ & $\ket{0}$ & $\ket{1}$ & $\Ket{\psi_2}=\frac{1}{\sqrt{2}}(\Ket{000}-\Ket{001}+\Ket{010}+\Ket{011}+\Ket{100}-\Ket{101}-\Ket{110}-\Ket{111})$ & 20 \\ \hline
$\ket{0}$ & $\ket{1}$ & $\ket{0}$ & $\Ket{\psi_3}=\frac{1}{\sqrt{2}}(\Ket{000}+\Ket{001}-\Ket{010}+\Ket{011}+\Ket{100}+\Ket{101}+\Ket{110}-\Ket{111})$ & 20 \\ \hline
$\ket{0}$ & $\ket{1}$ & $\ket{1}$ & $\Ket{\psi_4}=\frac{1}{\sqrt{2}}(\Ket{000}-\Ket{001}-\Ket{010}-\Ket{011}+\Ket{100}-\Ket{101}+\Ket{110}+\Ket{111})$ & 23 \\ \hline
$\ket{1}$ & $\ket{0}$ & $\ket{0}$ & $\Ket{\psi_5}=\frac{1}{\sqrt{2}}(\Ket{000}+\Ket{001}+\Ket{010}-\Ket{011}-\Ket{100}-\Ket{101}+\Ket{110}-\Ket{111})$ & 20 \\ \hline
$\ket{1}$ & $\ket{0}$ & $\ket{1}$ & $\Ket{\psi_6}=\frac{1}{\sqrt{2}}(\Ket{000}-\Ket{001}+\Ket{010}+\Ket{011}-\Ket{100}+\Ket{101}+\Ket{110}+\Ket{111})$ & 20 \\ \hline
$\ket{1}$ & $\ket{1}$ & $\ket{0}$ & $\Ket{\psi_7}=\frac{1}{\sqrt{2}}(\Ket{000}+\Ket{001}-\Ket{010}+\Ket{011}-\Ket{100}-\Ket{101}-\Ket{110}+\Ket{111})$ & 23 \\ \hline
$\ket{1}$ & $\ket{1}$ & $\ket{1}$ & $\Ket{\psi_8}=\frac{1}{\sqrt{2}}(\Ket{000}-\Ket{001}-\Ket{010}-\Ket{011}-\Ket{100}+\Ket{101}-\Ket{110}-\Ket{111})$ & 22 \\ \hline
\end{tabular}
\caption{Three-qubit cluster states after choosing different combinations of $\ket{a}$, $\ket{b}$ and $\ket{c}$. The last column contains the quantum cost for teleportation of a single qubit message via different three-qubit cluster states. \label{table:3q cluster States}}
\end{table*}

\begin{table*}[ht]
\centering
\renewcommand{\arraystretch}{1.75}
\begin{tabular}{|l|l|l|l|l|l|l|l|l|l|}
\hline
\multicolumn{1}{|c|}{\multirow{2}{*}{\textbf{\begin{tabular}[c]{@{}c@{}}Alice's Measurement \\ (Qubit \{1,2\})\end{tabular}}}} &
  \multicolumn{1}{c|}{\multirow{2}{*}{\textbf{\begin{tabular}[c]{@{}c@{}}Charlie Measurement \\ (Qubit \{3\})\end{tabular}}}} &
  \multicolumn{8}{c|}{\textbf{Bob's Operation (Qubit\{4\})}} \\ \cline{3-10} 
\multicolumn{1}{|c|}{} &
  \multicolumn{1}{c|}{} &
  $\ket{\psi_1}$ &
  $\ket{\psi_2}$ &
  $\ket{\psi_3}$ &
  $\ket{\psi_4}$ &
  $\ket{\psi_5}$ &
  $\ket{\psi_6}$ &
  $\ket{\psi_7}$ &
  $\ket{\psi_8}$ \\ \hline
\multirow{2}{*}{$\ket{00}$} &
  $\ket{0}$ &
  $\mathbb{I}$ &
  $\mathbb{Z}$ &
  $\mathbb{X}$ &
  $\mathbb{I}$ &
  $\mathbb{Z}$ &
  $\mathbb{ZX}$ &
  $\mathbb{ZX}$ &
  $\mathbb{Z}$ \\ \cline{2-10} 
 &
  $\ket{1}$ &
  $\mathbb{Z}$ &
  $\mathbb{ZX}$ &
  $\mathbb{ZX}$ &
  $\mathbb{Z}$ &
  $\mathbb{I}$ &
  $\mathbb{X}$ &
  $\mathbb{X}$ &
  $\mathbb{I}$ \\ \hline
\multirow{2}{*}{$\ket{01}$} &
  $\ket{0}$ &
  $\mathbb{X}$ &
  $\mathbb{I}$ &
  $\mathbb{I}$ &
  $\mathbb{X}$ &
  $\mathbb{ZX}$ &
  $\mathbb{Z}$ &
  $\mathbb{Z}$ &
  $\mathbb{ZX}$ \\ \cline{2-10} 
 &
  $\ket{1}$ &
  $\mathbb{ZX}$ &
  $\mathbb{Z}$ &
  $\mathbb{Z}$ &
  $\mathbb{ZX}$ &
  $\mathbb{X}$ &
  $\mathbb{I}$ &
  $\mathbb{I}$ &
  $\mathbb{X}$ \\ \hline
\multirow{2}{*}{$\ket{10}$} &
  $\ket{0}$ &
  $\mathbb{Z}$ &
  $\mathbb{ZX}$ &
  $\mathbb{ZX}$ &
  $\mathbb{Z}$ &
  $\mathbb{I}$ &
  $\mathbb{X}$ &
  $\mathbb{X}$ &
  $\mathbb{I}$ \\ \cline{2-10} 
 &
  $\ket{1}$ &
  $\mathbb{I}$ &
  $\mathbb{X}$ &
  $\mathbb{X}$ &
  $\mathbb{I}$ &
  $\mathbb{Z}$ &
  $\mathbb{ZX}$ &
  $\mathbb{ZX}$ &
  $\mathbb{Z}$ \\ \hline
\multirow{2}{*}{$\ket{11}$} &
  $\ket{0}$ &
  $\mathbb{ZX}$ &
  $\mathbb{Z}$ &
  $\mathbb{Z}$ &
  $\mathbb{ZX}$ &
  $\mathbb{X}$ &
  $\mathbb{I}$ &
  $\mathbb{I}$ &
  $\mathbb{X}$ \\ \cline{2-10} 
 &
  $\ket{1}$ &
  $\mathbb{X}$ &
  $\mathbb{I}$ &
  $\mathbb{I}$ &
  $\mathbb{X}$ &
  $\mathbb{ZX}$ &
  $\mathbb{Z}$ &
  $\mathbb{Z}$ &
  $\mathbb{ZX}$ \\ \hline
\end{tabular}
\caption{Controlled operations that Bob needs to apply to recover the message while teleporting via the three-qubit cluster state.}
\label{table:3q Cluster Operation}
\end{table*}


\subsection{Teleportation using Brown \emph{et al.} state } \label{Sec6}
\begin{figure}[!htb]
    \centering
    \includegraphics[width=0.5\textwidth]{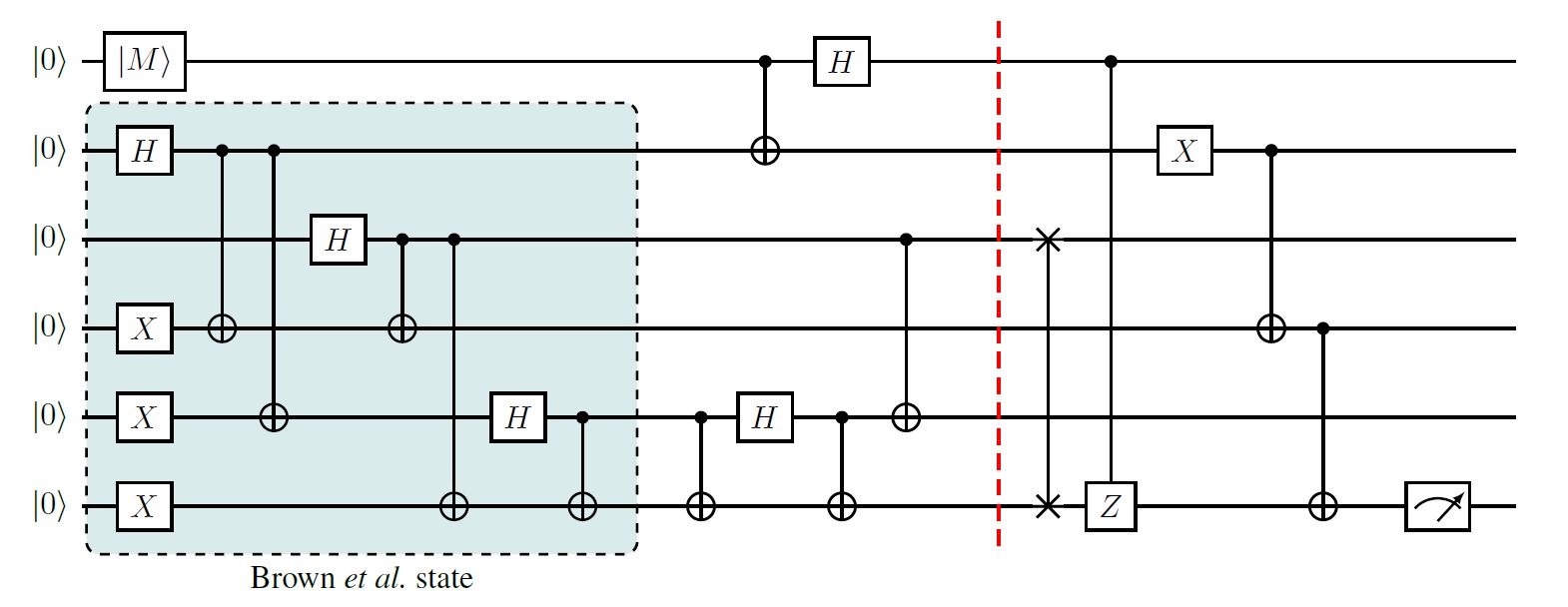}
    \caption{Teleportation protocol for teleporting a single-qubit message $\ket{M}= \alpha \ket{0} + \beta \ket{1}$ via Brown \emph{et al.} state, the controlled operations are represented after the red-dashed line.}
    \label{fig:Brown_Channel}
\end{figure}

Alice starts with a five-qubit entangled state called Brown \emph{et al.} state, which is considered to be a maximally entangled five-qubit state showing a high degree of entanglement \cite{Brown2005,anagha2020new}. It was verified by Borras \emph{et al.} \cite{Borras2007} that Brown \emph{et al.} state is highly efficient for teleportation. The state can be described in the following Eq. \eqref{eq:Brown}.
\begin{equation}
    \ket{\psi} = \frac{1}{2}[ \ket{001}\ket{\phi^{-}} + \ket{010}\ket{\psi^{-}} + \ket{100}\ket{\phi^{+}} +\ket{111}\ket{\psi^{+}}]
    \label{eq:Brown}
\end{equation}
where $\ket{\phi^{\pm}} = \frac{1}{\sqrt{2}} (\ket{01} \pm \ket{10})$ and $\ket{\psi^{\pm}} = \frac{1}{\sqrt{2}} (\ket{00} \pm \ket{11})$. Now, the teleportation of a single-qubit message $\ket{M} = (\alpha \ket{0} +\beta \ket{1})_1$ is performed using the Brown \emph{et al.} state, $\ket{\psi}_{23456}$. Thus, the combined state after taking the tensor product of message qubit $\ket{M}$ and the Brown \emph{et al.} state $\ket{\psi}$, the total state is given by Eq. \eqref{eq:Brown Tel},

\begin{eqnarray}
    \ket{\xi} &=& \ket{M}_1 \otimes \ket{\psi_{1}}_{2345}\nonumber \\ 
              &=& (\alpha \ket{0} + \beta \ket{1})_1 \otimes \Big( \frac{1}{2}[ \ket{001}\ket{\phi^{-}} + \ket{010}\ket{\psi^{-}} \nonumber \\  &+& \ket{100}\ket{\phi^{+}} +\ket{111}\ket{\psi^{+}}] \Big)_{2345}
\label{eq:Brown Tel}    
\end{eqnarray}
The teleportation protocol for teleporting the single-qubit message $\ket{M}=\alpha \ket{0}+\beta\ket{1}$ via the Brown \emph{et al.} state is given in the quantum circuit described in Fig. \ref{fig:Brown_Channel}. Alice applies the Bell basis measurement on the first and second qubit, i.e., $CX(1,2)$, $H(1)$ and the other operation described in Fig. \ref{fig:Brown_Channel} on the last four qubits, the final state is shown in Eq. \eqref{eq:Brown final_eq},

\begin{eqnarray}
&&\hspace*{-0.5cm} \ket{\xi} = \frac{\ket{00}_{12}}{4} \Big( \ket{011}_{345} (\alpha \ket{0} + \beta \ket{1})_6 + \ket{100}_{345} (\alpha \ket{1} \nonumber \\
    && + \beta \ket{0})_6 \Big) + \frac{\ket{01}_{12}}{4} \Big( \ket{011}_{345} (\alpha \ket{1} + \beta \ket{0})_6 \nonumber\\ 
    &&+ \ket{100}_{345} (\alpha \ket{0} \nonumber + \beta \ket{1})_6 \Big) +  \frac{\ket{10}_{12}}{4} \Big( \ket{011}_{345} (\alpha \ket{0} \nonumber \\ 
    &&- \beta \ket{1})_6 + \ket{100}_{345} (\alpha \ket{1} - \beta \ket{0})_6 \Big) + \frac{\ket{11}_{12}}{4} \nonumber \\
    && \Big( \ket{011}_{345} (\alpha \ket{1} - \beta \ket{0})_6 + \ket{100}_{345} (\alpha \ket{0} - \beta \ket{1})_6 \Big) \nonumber \\
\label{eq:Brown final_eq}    
\end{eqnarray}

After the message qubit $\ket{M}=\alpha \ket{0} + \beta \ket{1}$ is teleported to Bob's qubit, Bob needs to apply appropriate unitary operations on his qubit to recover the message successfully using the Brown \emph{et al.} state are shown in Table \ref{table:Brown control operations} and represented in the quantum circuit (Fig. \ref{fig:Brown_Channel}) after the red dashed line. The quantum cost for teleporting the single-qubit secret message via Brown \emph{et al.} state is 24.

\begin{table*}[]
\centering
\begin{tabular}{|l|l|l|l|}
\hline
\multicolumn{1}{|c|}{\textbf{\begin{tabular}[c]{@{}c@{}}Aice's Measurement\\ (Qubit\{1,2\})\end{tabular}}} & \multicolumn{1}{c|}{\textbf{Qubit\{3,4,5\}}} & \multicolumn{1}{c|}{\textbf{\begin{tabular}[c]{@{}c@{}}Last Qubit Measurement\\ Qubit\{6\}\end{tabular}}} & \multicolumn{1}{c|}{\textbf{Operations}} \\ \hline
\multirow{2}{*}{$\ket{00}_{12}$} & $\ket{110}_{345}$ & $\alpha \ket{0} + \beta \ket{1}$ & $\mathbb{I}$ \\ \cline{2-4} 
 & $\ket{011}_{345}$ & $\alpha \ket{1} + \beta \ket{0}$ & $\mathbb{X}$ \\ \hline
\multirow{2}{*}{$\ket{01}_{12}$} & $\ket{110}_{345}$ & $\alpha \ket{1} + \beta \ket{0}$ & $\mathbb{X}$ \\ \cline{2-4} 
 & $\ket{011}_{345}$ & $\alpha \ket{0} + \beta \ket{1}$ & $\mathbb{I}$ \\ \hline
\multirow{2}{*}{$\ket{10}_{12}$} & $\ket{110}_{345}$ & $\alpha \ket{0} - \beta \ket{1}$ & $\mathbb{Z}$ \\ \cline{2-4} 
 & $\ket{011}_{345}$ & $\alpha \ket{1} - \beta \ket{0}$ & $\mathbb{ZX}$ \\ \hline
\multirow{2}{*}{$\ket{11}_{12}$} & $\ket{110}_{345}$ & $\alpha \ket{1} - \beta \ket{0}$ & $\mathbb{ZX}$ \\ \cline{2-4} 
 & $\ket{011}_{345}$ & $\alpha \ket{0} - \beta \ket{1}$ & $\mathbb{Z}$ \\ \hline
\end{tabular}
\caption{Controlled operations that Bob needs to apply to recover the secret message during teleportation via Brown \emph{et al.} state.}
\label{table:Brown control operations}
\end{table*}

\subsection{Teleportation using Borras \emph{et al.}} state \label{Sec7}
Borras \emph{et al.} introduced a highly entangled six-qubit quantum state, which is not decomposable into product of Bell states \cite{Borras2007}. Borras \emph{et al.} state exhibits genuine entanglement according to many measures . 
Also, no other six-qubit pure state has been found that evolves to a mixed state with a higher amount of entanglement \cite{choudhury2009quantum}. The Borras \emph{et al.} state can be written in the terms of Bell basis given by Eq. \eqref{Borras_Eq1}.
\begin{eqnarray}
    \ket{\psi} &=& \frac{1}{4}\Big(\ket{000}(\ket{0}\ket{\psi^{+}} + \ket{1}\ket{\phi^{+}}) +\ket{001}(\ket{0}\ket{\phi^{-}} \nonumber\\ 
    &-& \ket{1}\ket{\psi^{-}})
    +\ket{010}(\ket{0}\ket{\phi^{+}} - \ket{1}\ket{\psi^{+}}) \nonumber\\ 
    &+& \ket{011}(\ket{0}\ket{\psi^{-}} + \ket{1}\ket{\phi^{-}}) + \ket{100}(-\ket{0}\ket{\phi^{-}} \nonumber\\ 
    &-& \ket{1}\ket{\psi^{-}}) +\ket{101}(-\ket{0}\ket{\psi^{+}} + \ket{1}\ket{\phi^{+}}) \nonumber\\ 
    &+&\ket{110}(\ket{0}\ket{\psi^{-}} - \ket{1}\ket{\phi^{-}}) \nonumber \\
    &+& \ket{111}(\ket{0}\ket{\phi^{+}} + \ket{1}\ket{\psi^{+}}) \Big) 
\label{Borras_Eq1}    
\end{eqnarray}

where $\ket{\phi^{\pm}} = \frac{1}{\sqrt{2}} (\ket{01} \pm \ket{10})$ and $\ket{\psi^{\pm}} = \frac{1}{\sqrt{2}} (\ket{00} \pm \ket{11})$. To teleport the single-qubit message $\ket{M} = \alpha \ket{0} + \beta \ket{1}$ via Borras \emph{et al.} state channel, the combined state after taking the tensor product of $\ket{M}$ and $\ket{\psi}$ given in Eq. \eqref{eq:Borras}.

\begin{eqnarray}
    \ket{\xi} &=& \ket{M}_1 \otimes {\ket{\psi}}_{234567} \nonumber\\  
    &=& (\alpha \ket{0} + \beta \ket{1})_1 \otimes \frac{1}{4} \Big( \ket{000}(\ket{0}\ket{\psi^{+}} + \ket{1}\ket{\phi^{+}}) \nonumber \\
    &+& \ket{001}(\ket{0}\ket{\phi^{-}} - \ket{1}\ket{\psi^{-}}) + \ket{010} (\ket{0}\ket{\phi^{+}} \nonumber \\
    &-& \ket{1}\ket{\psi^{+}}) \nonumber + \ket{011}(\ket{0}\ket{\psi^{-}} + \ket{1}\ket{\phi^{-}}) \nonumber \\ 
    &+&  \ket{100}(-\ket{0}\ket{\phi^{-}} - \ket{1}\ket{\psi^{-}}) +\ket{101}(-\ket{0}\ket{\psi^{+}} \nonumber \\
    &+& \ket{1}\ket{\phi^{+}}) + \ket{110}(\ket{0}\ket{\psi^{-}} - \ket{1}\ket{\phi^{-}}) \nonumber \\ 
    &+& \ket{111}(\ket{0}\ket{\phi^{+}} + \ket{1}\ket{\psi^{+}}) \Big)_{234567} 
\label{eq:Borras}    
\end{eqnarray}

\begin{figure}[!htb]
    \centering
    \includegraphics[width=0.45\textwidth]{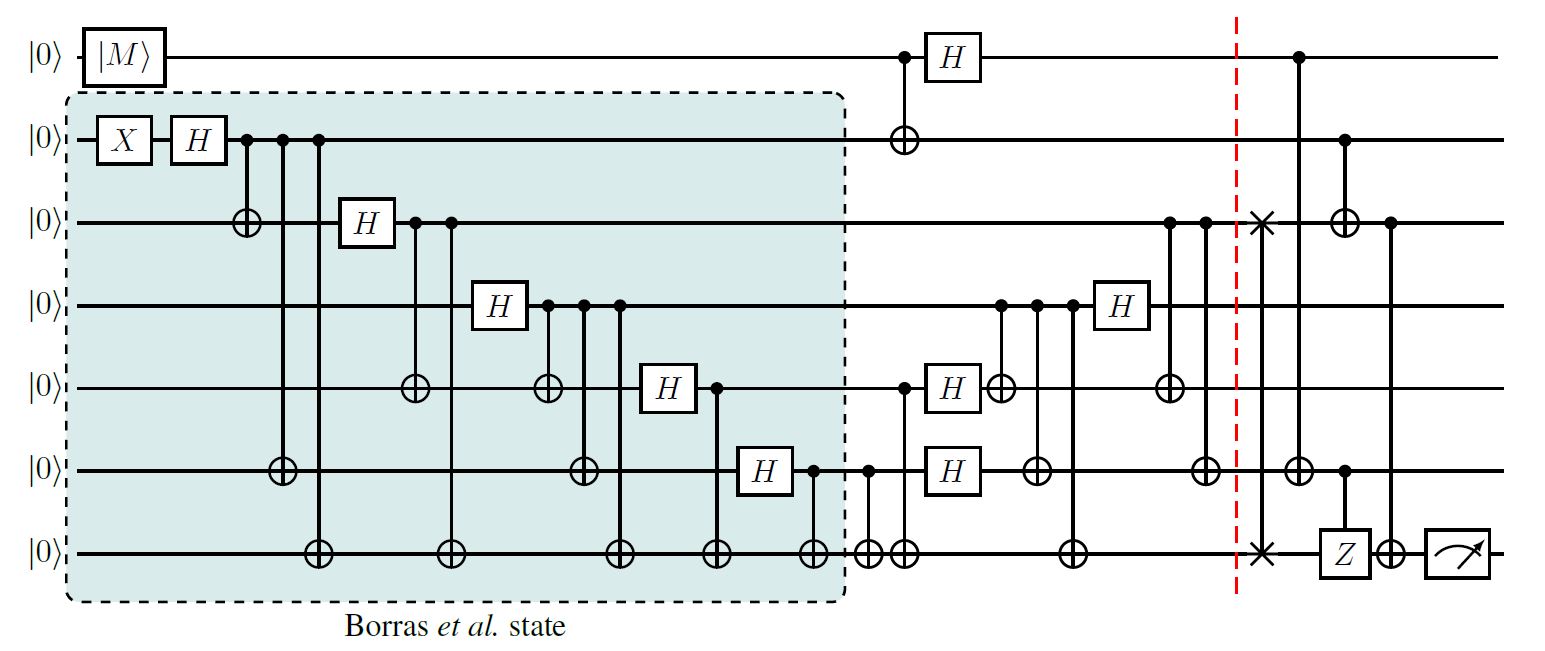}
    \caption{Teleportation protocol of a single-qubit message $\ket{M}= \alpha \ket{0} + \beta \ket{1}$ using Borras \emph{et al.} state,  the controlled operations are represented after the red-dashed line.}
    \label{fig_BorrasChannel}
\end{figure}

Now, Alice applies the Bell basis measurement on the first and second qubit, i.e. $CX(1,2)$, $H(1)$ and other operations on the last five qubits depicted in quantum circuit (Fig. \ref{fig_BorrasChannel}), the final state of the system is given in Eq. \eqref{eq:Borras_final},

\begin{eqnarray}
&&  \ket{\xi} = 
    \frac{\ket{00}_{12}}{2} \Big( \ket{0000}_{3456} (\alpha \ket{0} + \beta \ket{1})_7 + \ket{1001}_{3456} \nonumber \\
    && (\alpha \ket{1} - \beta \ket{0})_7 \Big) + 
    \frac{\ket{01}_{12}}{4} \Big( \ket{0000}_{3456} (\alpha \ket{1} + \beta \ket{0})_7 \nonumber \\ 
    &&+ \ket{1001}_{3456} (-\alpha \ket{0} + \beta \ket{1})_7 \Big) +
    \frac{\ket{10}_{12}}{4} \Big( \ket{0000}_{3456} \nonumber \\ 
    && (\alpha \ket{0} - \beta \ket{1})_7 + \ket{1001}_{3456} (\alpha \ket{1} + \beta \ket{0})_7 \Big) \nonumber \\
    &&+ \frac{\ket{11}_{12}}{4} \Big( \ket{0000}_{3456} (\alpha \ket{1} - \beta \ket{0})_7 + \ket{1001}_{3456} \nonumber \\ && (\alpha \ket{0} + \beta \ket{1})_7 \Big)
\label{eq:Borras_final}    
\end{eqnarray}

After reaching this state, Bob applies proper unitary operations on the seventh qubit to recover the teleported message. The unitary operations are shown in Table \ref{table:Borras control operations}. The teleportation protocol is given in the following quantum circuit (Fig. \ref{fig_BorrasChannel}). The quantum cost for teleporting the single-qubit message via Borras \emph{et al.} state is 38.

\begin{table*}
\centering
\begin{tabular}{|l|l|l|l|}
\hline
\multicolumn{1}{|c|}{\textbf{\begin{tabular}[c]{@{}c@{}}Aice's Measurement\\ (Qubit\{1,2\})\end{tabular}}} & \multicolumn{1}{c|}{\textbf{Qubit\{3,4,5,6\}}} & \multicolumn{1}{c|}{\textbf{\begin{tabular}[c]{@{}c@{}}Last Qubit Measurement\\ Qubit\{7\}\end{tabular}}} & \multicolumn{1}{c|}{\textbf{Operations}} \\ \hline
\multirow{2}{*}{$\ket{00}_{12}$} & $\ket{0000}_{3456}$ & $\alpha \ket{0} + \beta \ket{1}$ & $\mathbb{I}$ \\ \cline{2-4} 
 & $\ket{0011}_{3456}$ & $\alpha \ket{1} - \beta \ket{0}$ & $\mathbb{ZX}$ \\ \hline
\multirow{2}{*}{$\ket{01}_{12}$} & $\ket{0000}_{3456}$ & $\alpha \ket{1} + \beta \ket{0}$ & $\mathbb{X}$ \\ \cline{2-4} 
 & $\ket{0011}_{3456}$ & $-\alpha \ket{0} + \beta \ket{1}$ & $\mathbb{Z}$ \\ \hline
\multirow{2}{*}{$\ket{10}_{12}$} & $\ket{0000}_{3456}$ & $\alpha \ket{0} - \beta \ket{1}$ & $\mathbb{Z}$ \\ \cline{2-4} 
 & $\ket{0011}_{3456}$ & $\alpha \ket{1} + \beta \ket{0}$ & $\mathbb{X}$ \\ \hline
\multirow{2}{*}{$\ket{11}_{12}$} & $\ket{0000}_{3456}$ & $\alpha \ket{1} - \beta \ket{0}$ & $\mathbb{ZX}$ \\ \cline{2-4} 
 & $\ket{0011}_{3456}$ & $\alpha \ket{0} + \beta \ket{1}$ & $\mathbb{I}$ \\ \hline
\end{tabular}
\caption{Controlled operations that Bob needs to apply to recover the teleported message via Borras \emph{et al.} state.}
\label{table:Borras control operations}
\end{table*}

\begin{figure*}[!htb]
     \centering
     \subfigure{\includegraphics[width=0.3\textwidth]{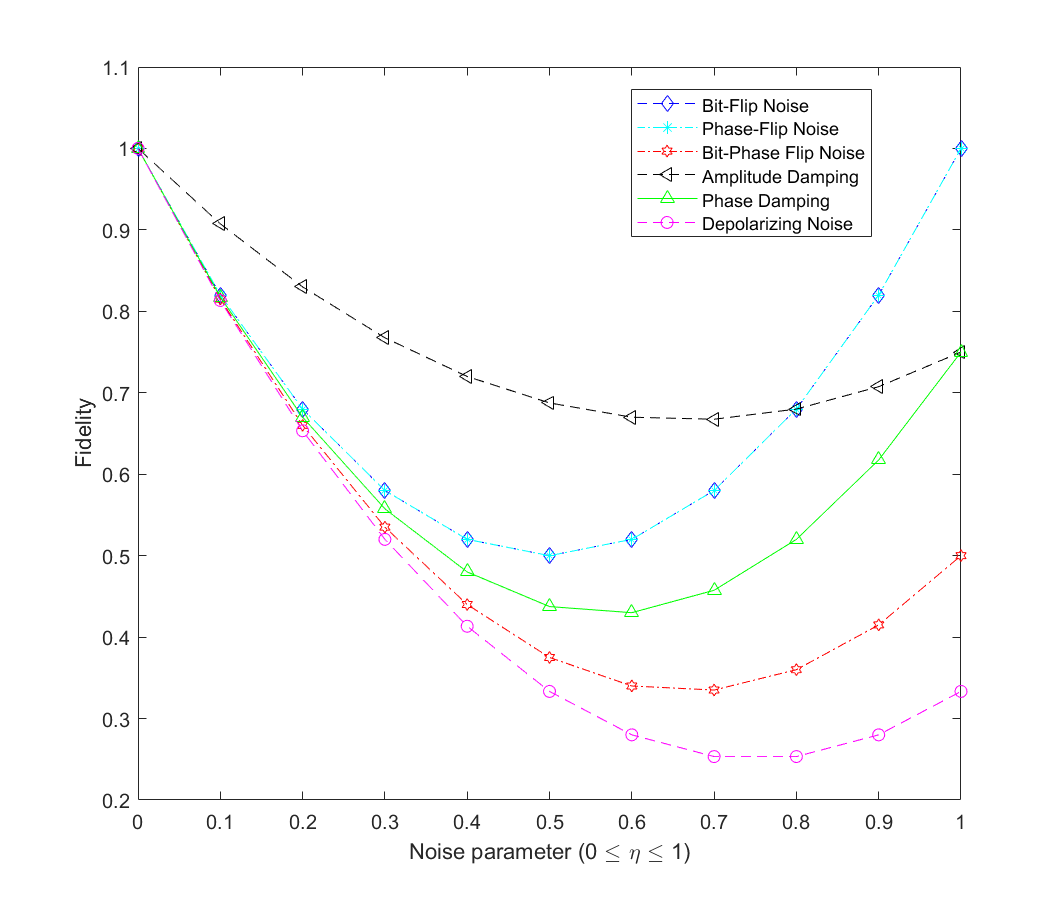}} \label{fig:Graph Plot 1}
     \subfigure{\includegraphics[width=0.3\textwidth]{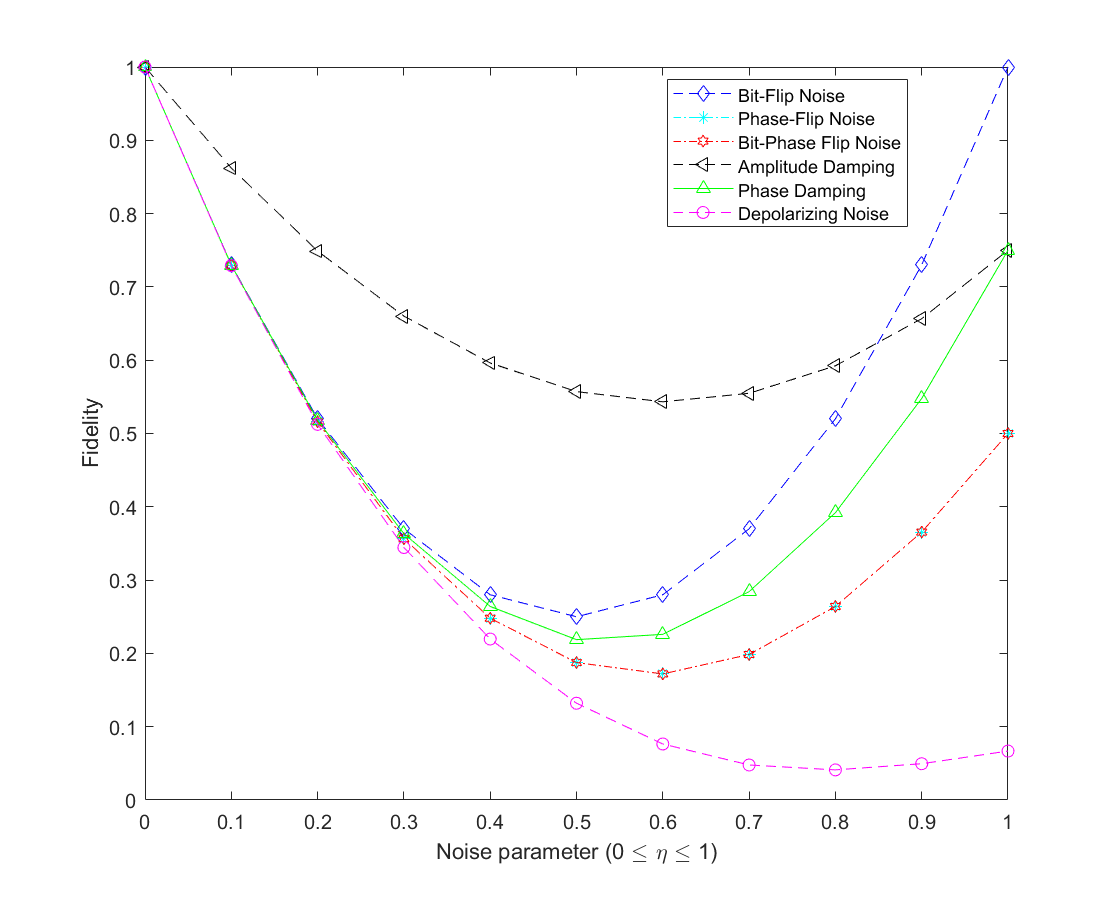}} \label{fig:Graph Plot 2}
     \subfigure{\includegraphics[width=0.3\textwidth]{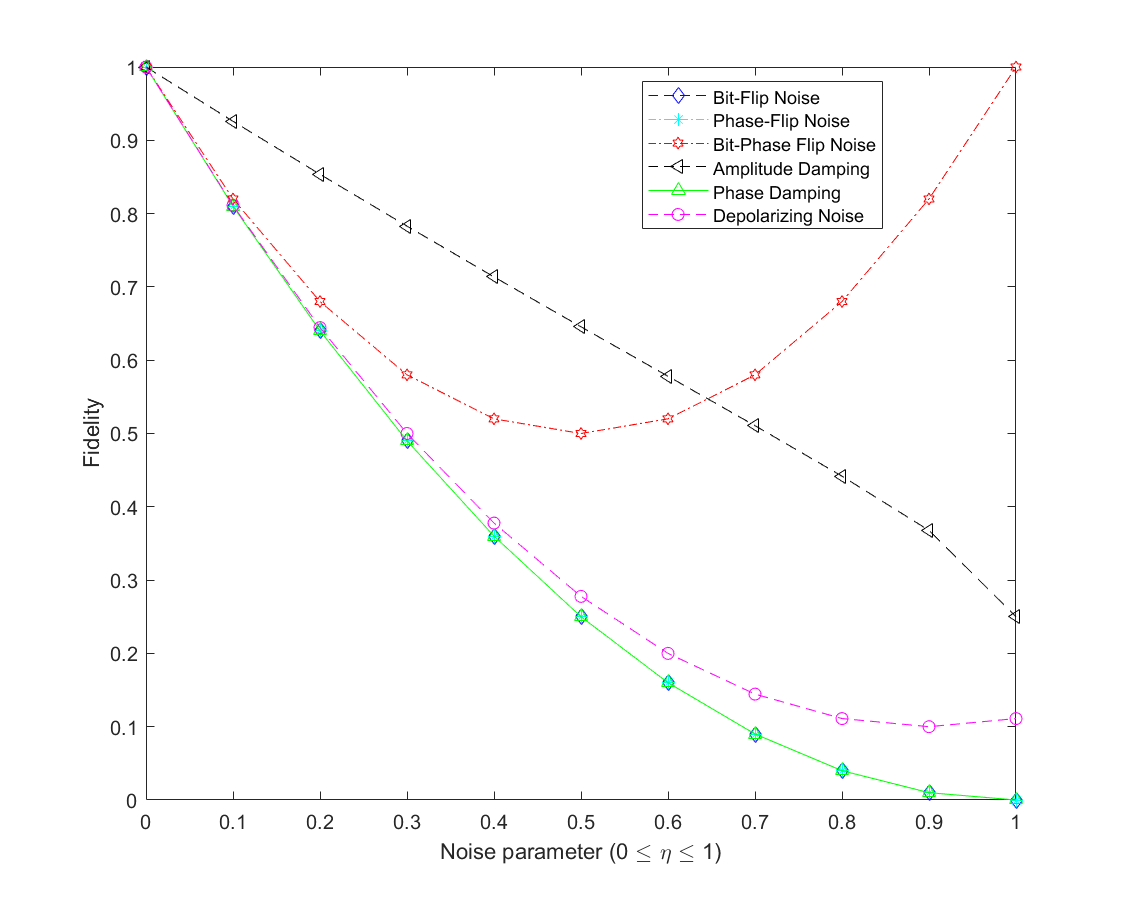}} \label{fig:Graph Plot 3}
     \subfigure{\includegraphics[width=0.3\textwidth]{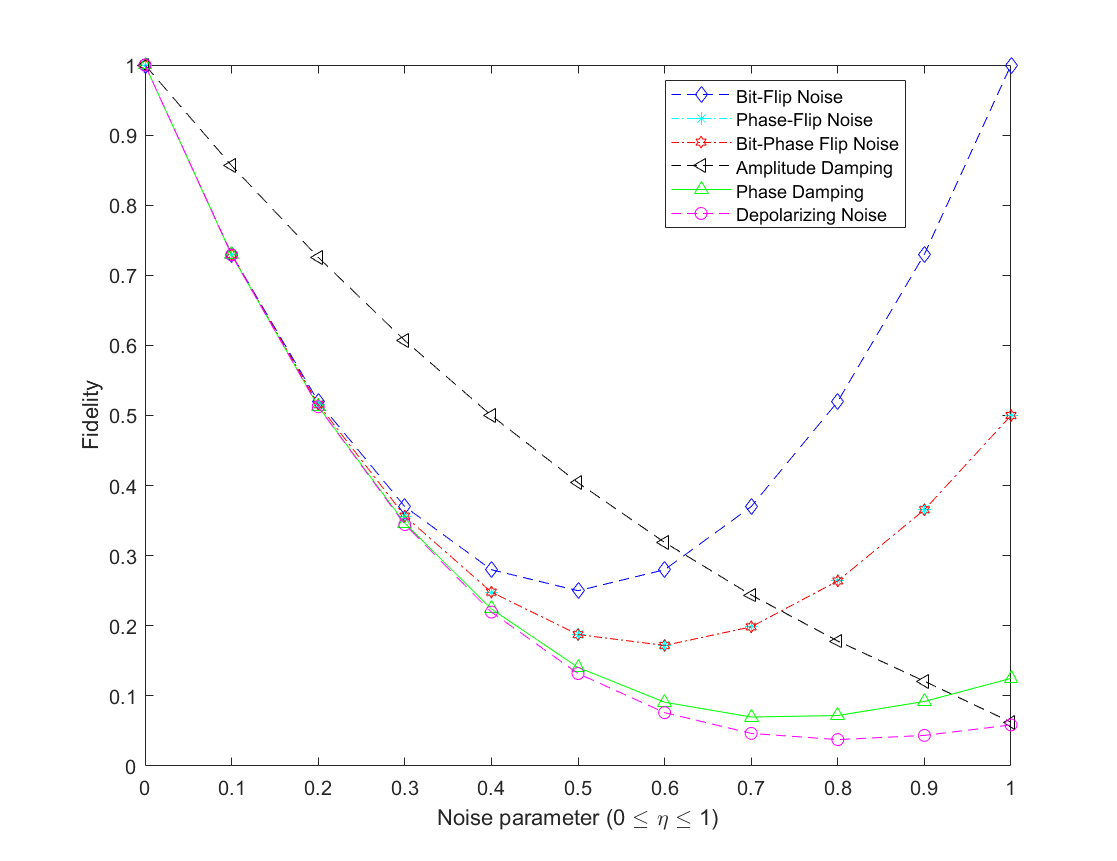}} \label{fig:Graph Plot 4}
     \subfigure{\includegraphics[width=0.3\textwidth]{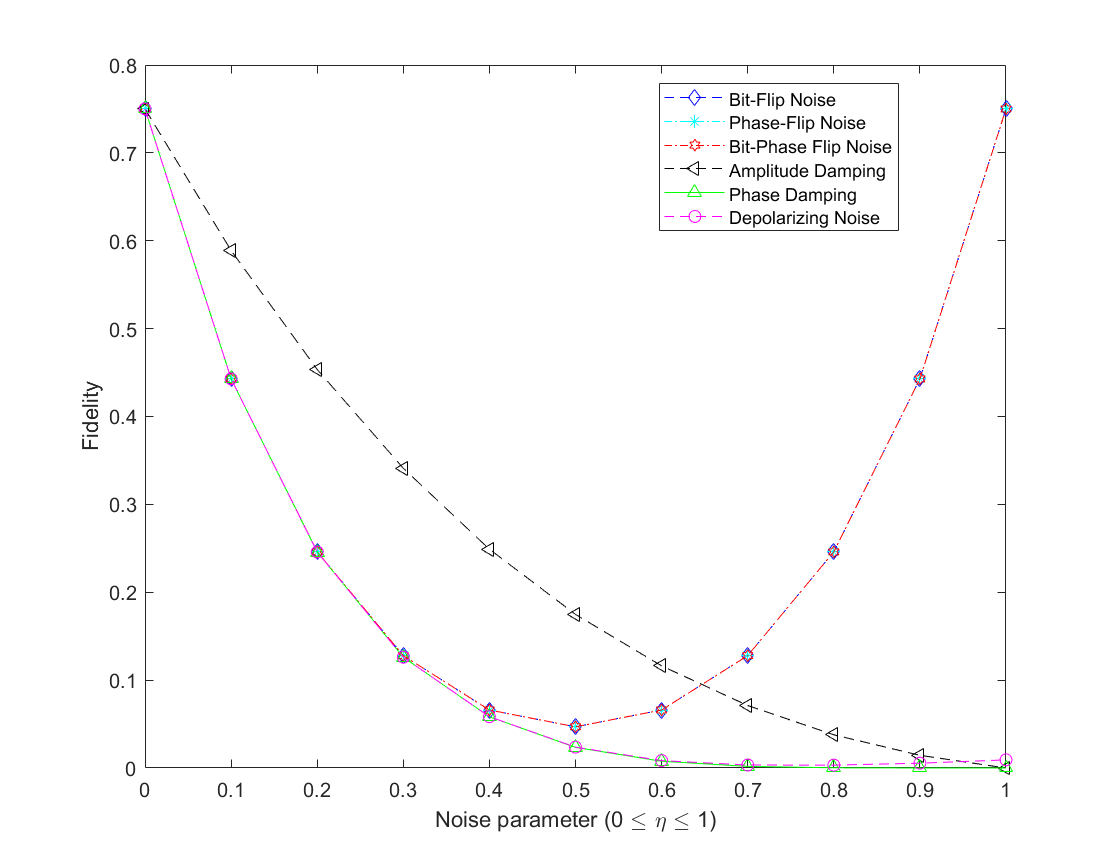}} \label{fig:Graph Plot 5}
     \subfigure{\includegraphics[width=0.3\textwidth]{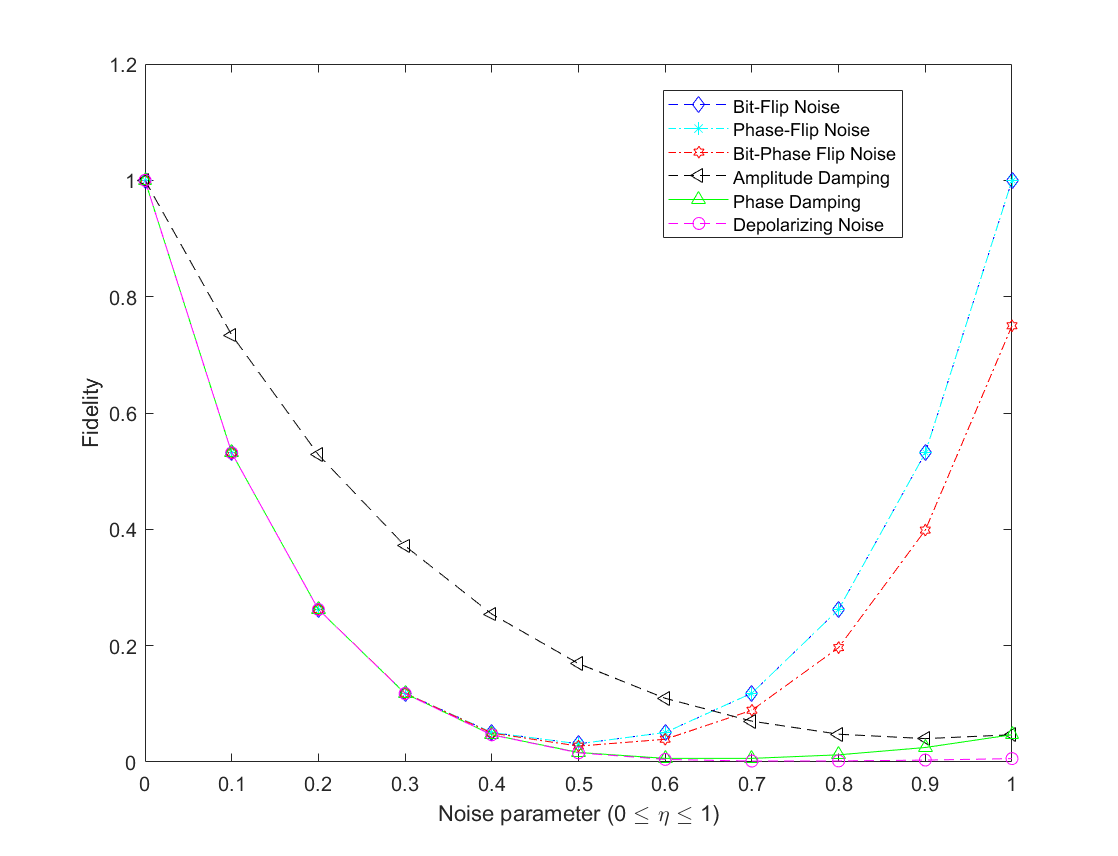}} \label{fig:Graph Plot 6}
     \caption{The effect of all six noses on the teleportation is visualized through a graphical representation of variation of fidelity against the noise parameter $\eta$. Here ($\eta \in \{\eta_B, \eta_W, \eta_F, \eta_A, \eta_P, \eta_D \} $ for bit-flip noise, phase flip noise, bit-phase flip noise, amplitude damping, phase damping and depolarizing noise respectively). The teleportation channels are: (a) Bell channel (b) GHZ channel (c) Two-qubit cluster state channel (d) Three-qubit cluster state channel (e) Brown \emph{et al.} channel (f) Borras \emph{et al.} channel.} 
     \label{fig:Graph Plot}
 \end{figure*}


\section{Effect of noisy environment on the quantum channel }\label{Sec8}
Practically, it is impossible to perform quantum teleportation without noise in the entangled channel. That means noise is an unavoidable feature of real experiments. So, the best we can do is to study the sources and effects of these noises in our system and minimize them as much as possible. Here, we conveniently study the impact of six different kinds of noises present in all the teleportation protocols performed above. We do this by checking the evolution of density matrix $\rho=\ket{\psi}\bra{\psi}$ under the effect of Kraus operators. The six noise models we study here are the bit-flip noise, phase-flip noise, bit-phase-flip noise, amplitude damping, phase damping, and depolarising noise. Using the operator sum representation, the interaction of noise in the quantum channel can be represented with the help of Kraus operators $E_k$. The action of noise on a particular qubit $k$ described by the density matrix $\rho_k$ is given by the Eq. \eqref{eq:Xi}
\begin{equation}
    \xi^r(\rho_k) = \sum_{j=1}^{n} \big({E_j}\big)\ \rho_k\ \big({E_{j}}\big)^{\dagger}
\label{eq:Xi}
\end{equation}
where $r \in \{B,W,F,A,P,D\}$ for bit-flip, phase-flip, bit-phase-flip, amplitude damping, phase damping and depolarizing damping respectively, $m \in \{0,1\}$ for $r=B,W,F,A$, $m \in \{0,1,2\}$ for $r=P$ and $m \in \{0,1,2,3\}$ for $r=D$. In the presence of noisy environment, the shared entangled state would become a mixed state after the distribution of the qubits. To recover the original message, Bob has to make appropriate unitary operations on his own qubit. The final state $\rho_{out} ^r$ can be expressed as the density matrix shown in the following Eq. \eqref{eq:rho_out}.

\begin{equation} 
    \rho_{out}^r = Tr_{i_1i_2...i_{n-1}} \{ \mathbb{U} [\rho_k \otimes  \xi^r(\rho_l)] \mathbb{U}^{\dagger} \}
    \label{eq:rho_out}
\end{equation}
where $Tr_{i_1i_2...i_{n-1}}$ is the partial trace over the qubits $i_1,i_2,...,i_{n-1}$ and $\mathbb{U}$ is the unitary operation to describe the teleportation process, which is given by the following Eq. \eqref{eq:U},

\begin{eqnarray}
\mathbb{U} &=& \{\mathbb{I}_1 \otimes \mathbb{I}_2 \otimes . . .\mathbb{I}_{n-1} \otimes \sigma_{n} ^{i_1 i_2 ... i_{n-1}} \} \nonumber \\
&& \{\ket{\phi}_{12} \bra{\phi}_{12} \otimes \mathbb{I}_3 ... \otimes \mathbb{I}_n \} \{\mathbb{I}_1 \otimes \mathbb{I}_2 ... U_{j_1 j_2} ... \otimes \mathbb{I}_n \} \nonumber \\
&& \{\mathbb{I}_1 \otimes \mathbb{I}_2 ... U_{k_1} ... \otimes \mathbb{I}_n \}
\label{eq:U}
\end{eqnarray}

where $\sigma_{n} ^{i_1 i_2 ... i_{n-1}}$ is the the Bob's recovery unitary operations after Alice has measured her qubit and communicated her result to Bob via a classical channel, $\ket{\phi}_{12} \bra{\phi}_{12}$ is the Bell basis measurement on the first two qubits, $U_{j_1 j_2}$ represents the CNOT gate from qubit $j_1$ to $j_2$ and $U_{k_1}$ represents the unitary gate on qubit $k_1$. Depending upon the different choices of the entangled channel, these unitary operations change. Now, the effect of noise in the entangled channel can be visualized by calculating the fidelity between the initial single qubit message $\ket{\psi}$ and the density matrix $\rho_{out}^r$. Fidelity represents the closeness of the two quantum states, and it gives a mathematical prescription for the quantification of the degree of similarity of a pair of quantum states. The mathematical expression is given by Eq. \eqref{eq:Fid} \cite{liang2019quantum}. 
 \begin{equation}
     F = \bra{\Psi} \rho_{out}^r \ket{\Psi}
     \label{eq:Fid}
 \end{equation}

Some quantum state information will be lost during the teleportation through a noisy environment, so fidelity is a perfect metric to measure how much information is lost. If the fidelity $F=1$, then this indicated the ideal case where no information is lost, and the teleportation is perfect. Meanwhile, $F=0$ implies that all the information is lost, and the quantum states before and after the teleportation are orthogonal. Thus, the fidelity ranges between $0$ and $1$, i.e., $0\leq F \leq 1$. We now discuss the effect of six types of noises (bit-flip, phase-flip, bit-phase-flip, amplitude damping, phase damping and depolarizing noise) in the next section.


\subsection{Bit-flip noisy environment}
In the presence of bit-flip noise, the quantum state $\ket{0}$ is changed to $\ket{1}$ and vice-versa with probability $\eta_B$ and the qubits remain unchanged with the probability $(1-\eta_B)$\cite{fortes2015fighting, oh2002fidelity}. Its operations on a qubit can be described by Kraus operators given by the following matrices in Eq.\eqref{eqn_bit_noise}  

\begin{eqnarray}
 && E_0^B= \sqrt{1-\eta_B} \mathbb{I} =
 \sqrt{1-\eta_B} \begin{pmatrix}
    1 & 0 \\
    0 & 1
  \end{pmatrix}, \nonumber\\ 
 && E_1^B= \sqrt{\eta_B}\mathbb{X} = \sqrt{\eta_B}
  \begin{pmatrix}
    0 & 1 \\
    1 & 0
  \end{pmatrix}
\label{eqn_bit_noise}
\end{eqnarray}

Where $0\leq \eta_{B} \leq 1$ represents the bit-flip error probability, which describes the probability of occurring error in the quantum state due to travel qubit, the effect of bit-flip noise on the six entangled channels can be seen by evaluating the affected density matrix after the introduction of noise, the affected density matrix under the bit-flip noise for all the six channels the Bell channel, GHZ channel, two-qubit cluster state, three-qubit cluster state, Brown \emph{et al.} state and the Borras \emph{et al.} are denoted by $\mathcal{E}_1^B (\rho), \mathcal{E}_2^B (\rho), \mathcal{E}_3^B (\rho), \mathcal{E}_4^B (\rho), \mathcal{E}_5^B (\rho), \mathcal{E}_6^B (\rho)$ respectively, given in Eq. \eqref{eq:bit_1} to \eqref{eq:bit_6}.

\begin{widetext}
The affected density matrices are influenced by the bit-flip noise are given by the following equations from Eq. \eqref{eq:bit_1} to Eq. \eqref{eq:bit_6}. Here $\ket{\Psi_6}$ is the Borras \emph{et al.} state given in Eq. \eqref{Borras_Eq1}
\begin{eqnarray}
&& \mathcal{E}_1^B (\rho) = \frac{1}{2} \Big\{ [(1-\eta_B)^2  + (\eta_B)^2 ][\ket{00} + \ket{11}][\bra{00} + \bra{11} ] \Big\}
\label{eq:bit_1} \\ 
&& \mathcal{E}_2^B (\rho) = \frac{1}{2} \Big\{[(1-\eta_B)^{3} + (\eta_B)^3][\ket{000} + \ket{111}][\bra{000} + \bra{111} ]\Big\} \label{eq:bit_2} \\
&& \mathcal{E}_3^B (\rho) = \frac{1}{4} \Big\{ [(1-\eta_B)^{2} ][\ket{00} +\ket{01} +\ket{10} - \ket{11}][\bra{00} + \bra{01} + \bra{10} - \bra{11} ] + (\eta_B)^2 [\ket{11} + \ket{01} + \ket{10}  - \ket{00}] \nonumber\\ 
    && ~~~~~~~~~~~~~~~~~~~~[\bra{11} + \bra{01} + \bra{10} - \bra{00} ] \Big\} \label{eq:bit_3} \\
&& \mathcal{E}_4^B (\rho) = \frac{1}{8} \Big\{ [(1-\eta_B)^{3} ][\ket{000} +\ket{001} +\ket{010} - \ket{011} + \ket{100} +\ket{101} - \ket{110} + \ket{111}][\bra{000} + \bra{001} + \bra{010} \nonumber \\ 
    &&~~~~~~~~~~~~~~~~ - \bra{011} + \bra{100} + \bra{101} - \bra{110} + \bra{111}] + (\eta_B)^3 [\ket{111} +\ket{110} +\ket{101} - \ket{100} + \ket{011} +\ket{010} - \ket{001} \nonumber \\
    &&~~~~~~~~~~~~~~~~ + \ket{000}][\bra{111} + \bra{110} + \bra{101} - \bra{100} + \bra{011} + \bra{010} - \bra{001} + \bra{000}] \Big\} \label{eq:bit_4} \\
&& \mathcal{E}_5^B (\rho) = \frac{1}{8} \Big\{[(1-\eta_B)^{5} ][-\ket{00101} + + \ket{00111} + \ket{01000} - \ket{01010} + \ket{10001} + \ket{10011} + \ket{11100} + \ket{11110}] \nonumber \\  
    &&~~~~~~~~~~~~~~~~~~~~ [-\bra{00101} + + \bra{00111} + \bra{01000} - \bra{01010} + \bra{10001} + \bra{10011} + \bra{11100} + \bra{11110}]  + (\eta_B)^5 [-\ket{11010}   \nonumber \\
     &&~~~~~~~~~~~~~~~~ +  \ket{11000} + \ket{101111} - \ket{10101} + \ket{01110} + \ket{01100} + \ket{00011} + \ket{00001}] [-\bra{11010} + + \bra{11000} + \bra{101111} \nonumber \\ 
     &&~~~~~~~~~~~~~~~~ - \bra{10101} + \bra{01110} + \bra{01100} + \bra{00011} + \bra{00001}]\Big\} \\ \label{eq:bit_5}
&& \mathcal{E}_6^B (\rho) = \frac{1}{32} \Big\{ [(1-\eta_B)^{6} ][\ket{\Psi_6}][\bra{\Psi_6}] 
    + (\eta_B)^6
    [\ket{111111} +
    \ket{000000} +
    \ket{111100} +
    \ket{000011} +
    \ket{111010} +
    \ket{000101} \nonumber \\ &&~~~~~~~~~~~~~~~~~+
    \ket{111001} +
    \ket{000110} +
    \ket{110110} +
    \ket{001001} +
    \ket{110000} +
    \ket{001111} +
    \ket{101110} +
    \ket{010001} +
    \ket{101010} \nonumber  \\ &&~~~~~~~~~~~~~~~~~~~~+
    \ket{010010} +
    \ket{100111} +
    \ket{011000} +
    \ket{100010} +
    \ket{011101} -
    \ket{110101} -
    \ket{001010} -
    \ket{110011} -
    \ket{001100} \nonumber \\  &&~~~~~~~~~~~~~~~~~~~~ -
    \ket{101011} -
    \ket{010100} - 
    \ket{101000} -
    \ket{010111} -
    \ket{100100} -
    \ket{011011} -
    \ket{100001} -
    \ket{011110}]
    [\bra{111111} \nonumber  \\ &&~~~~~~~~~~~~~~~~~~~~ +
    \bra{000000} +
    \bra{111100} +
    \bra{000011} +
    \bra{111010} +
    \bra{000101} + 
    \bra{111001} +
    \bra{000110} +
    \bra{110110} +
    \bra{001001} \nonumber  \\ &&~~~~~~~~~~~~~~~~~~~~ +
    \bra{110000} +
    \bra{001111} +
    \bra{101110} +
    \bra{010001} +
    \bra{101010} +
    \bra{010010} +
    \bra{100111} +
    \bra{011000} +
    \bra{100010}  \nonumber  \\ &&~~~~~~~~~~~~~~~~~~~~ +
    \bra{011101} -
    \bra{110101} -
    \bra{001010} -
    \bra{110011} -
    \bra{001100} -
    \bra{101011} -
    \bra{010100} -
    \bra{101000} -
    \bra{010111}  \nonumber \\ && ~~~~~~~~~~~~~~~~~~~~ -
    \bra{100100} -
    \bra{011011} -
    \bra{100001} -
    \bra{011110}] \Big\}
    \label{eq:bit_6}
\end{eqnarray}
\end{widetext}


\subsection{Phase-flip noisy environment}
In the presence of phase-flip noise, the phase of the qubit changes from $\ket{1}$ to $-\ket{1}$, it remains unchanged if the qubit is $\ket{0}$. It's Kraus operators \cite{fortes2015fighting, oh2002fidelity} are given by Eq. \eqref{eqn_phase_noise},

\begin{eqnarray}
  && E_0^W= \sqrt{1-\eta_W} \mathbb{I} =
 \sqrt{1-\eta_W} \begin{pmatrix}
    1 & 0 \\
    0 & 1
  \end{pmatrix}~, \nonumber\\  
  && E_1^W= \sqrt{\eta_W}\mathbb{Z} = \sqrt{\eta_W}
  \begin{pmatrix}
    1 & 0 \\
    0 & -1
  \end{pmatrix}
\label{eqn_phase_noise}
\end{eqnarray}
where $0\leq \eta_{P} \leq 1$ represents the phase-flip error probability, which describes the possibility of occurring error in the quantum state due to travel qubit. The effect of phase-flip noise on the the six entangled channels can be seen by evaluating the affected density matrix after noise has been introduced in the channel, the affected density matrix under the phase-flip noise for all the six channels the Bell channel, GHZ channel, two-qubit cluster state, three-qubit cluster state, Brown \emph{et al.} state and the Borras \emph{et al.} are denoted by $\mathcal{E}_1^W (\rho), \mathcal{E}_2^W (\rho), \mathcal{E}_3^W (\rho), \mathcal{E}_4^w (\rho), \mathcal{E}_5^W (\rho), \mathcal{E}_6^W (\rho)$ respectively, given in Eq. \eqref{eq:phase_1} to \eqref{eq:phase_6}.


\begin{widetext}
The affected density matrices are influenced by the phase-flip noise are given by the following equations from Eq. \eqref{eq:phase_1} to Eq. \eqref{eq:phase_6}. Here $\ket{\Psi_6}$ is the Borras \emph{et al.} state given in Eq. \eqref{Borras_Eq1}.
\begin{eqnarray}
   && \mathcal{E}_1^W (\rho) = \frac{1}{2} \Big\{[(1-\eta_W)^2 + \eta_W^2][\ket{00} +  \ket{11}][\bra{00} + \bra{11}] \Big\} \label{eq:phase_1} \\
    && \mathcal{E}_2^W (\rho) = \frac{1}{2} \Big\{ (1-\eta_W)^3[\ket{000} + \ket{111}][\bra{000} + \bra{111}] + (\eta_W)^3[\ket{00} - \ket{11} ][\bra{00} - \bra{11} ] \Big\} \label{eq:phase_2} \\
    && \mathcal{E}_3^W (\rho) = \frac{1}{4} \Big\{ [(1-\eta_W)^{2} ][\ket{00} +\ket{01} +\ket{10} -\ket{11}][\bra{00} + \bra{01} + \bra{10} - \bra{11} ] + (\eta_W)^2 [\ket{00} - \ket{01} - \ket{10} - \ket{11}] \nonumber \\ 
    &&~~~~~~~~~~~~~~~~~~~~ [\bra{00} - \bra{01} - \bra{10} - \bra{11} ] \Big\} \label{eq:phase_3} \\
    && \mathcal{E}_4^W (\rho) = \frac{1}{8} \Big\{ [(1-\eta_W)^{3} ][\ket{000} +\ket{001} +\ket{010} - \ket{011} + \ket{100} +\ket{101} - \ket{110} + \ket{111}][\bra{000} + \bra{001} + \bra{010} \nonumber \\ 
    &&~~~~~~~~~~~~~~~~~ - \bra{011} + \bra{100} + \bra{101} - \bra{110} + \bra{111}] 
    + (\eta_W)^3 [\ket{000} -\ket{001} -\ket{010} - \ket{011} - \ket{100} + \ket{101} - \ket{110} \nonumber \\ 
    &&~~~~~~~~~~~~~~~~~ - \ket{111}][\bra{000} -\bra{001} -\bra{010} - \bra{011} - \bra{100} + \bra{101} - \bra{110} - \bra{111}] \Big\} \label{eq:phase_4} \\
    &&\mathcal{E}_5^W (\rho) = \frac{1}{2} \Big\{[(1-\eta_W)^{5} ][-\ket{00101} + \ket{00111} + \ket{01000} - \ket{01010} + \ket{10001} + \ket{10011} + \ket{11100} + \ket{11110}] \nonumber \\  
    && ~~~~~~~~~~~~~~~~~~~~ [-\bra{00101} +  \bra{00111} + \bra{01000} - \bra{01010} + \bra{10001} + \bra{10011} + \bra{11100} + \bra{11110}]  + (\eta_W)^5 [-\ket{00101} \nonumber \\ 
    &&~~~~~~~~~~~~~~~~~ - \ket{00111} - \ket{01000} - \ket{01010} + \ket{10001} - \ket{10011} - \ket{11100} + \ket{11110}]  [-\bra{00101} - \bra{00111} \nonumber \\  
    &&~~~~~~~~~~~~~~~~~ - \bra{01000} - \bra{01010} + \bra{10001} - \bra{10011} - \bra{11100} + \bra{11110}] \Big\} \label{eq:phase_5} \\ 
    && \mathcal{E}_6^W (\rho) = \frac{1}{32} \Big\{ (1-\eta_W)^{6} [\ket{\Psi_6}][\bra{\Psi_6}]
    + (\eta_W)^6 
    [\ket{000000} +
    \ket{111111} + 
    \ket{000011} + 
    \ket{111100} + 
    \ket{000101} + 
    \ket{111010}  \nonumber \\  &&~~~~~~~~~~~~~~~~~ + 
    \ket{000110} +
    \ket{111001} +
    \ket{001001} +
    \ket{110110} +
    \ket{001111} +
    \ket{110000} +
    \ket{010001} +
    \ket{101110} +
    \ket{010010} \nonumber \\  && ~~~~~~~~~~~~~~~~~~~~ + 
    \ket{101101} +
    \ket{011000} +
    \ket{100111} +
    \ket{011101} +
    \ket{100010} -
    \ket{001010} -
    \ket{110101} -
    \ket{001100} -
    \ket{110011} \nonumber \\  && ~~~~~~~~~~~~~~~~~~~~ -
    \ket{010100} - 
    \ket{101011} -
    \ket{010111} -
    \ket{101000} -
    \ket{011011} -
    \ket{100100} -
    \ket{011110} -
    \ket{100001}]
    [\bra{000000} \nonumber \\  &&~~~~~~~~~~~~~~~~~~~~ +
    \bra{111111} + 
    \bra{000011} + 
    \bra{111100} +
    \bra{000101} +
    \bra{111010} +
    \bra{000110} +
    \bra{111001} +
    \bra{001001} +
    \bra{110110} \nonumber \\  &&~~~~~~~~~~~~~~~~~~~~ +
    \bra{001111} +
    \bra{110000} +
    \bra{010001} +
    \bra{101110} +
    \bra{010010} +
    \bra{101101} +
    \bra{011000} +
    \bra{100111} +
    \bra{011101} \nonumber \\  &&~~~~~~~~~~~~~~~~~~~~ +
    \bra{100010} +
    \bra{001010} -
    \bra{110101} -
    \bra{001100} - 
    \bra{110011} - 
    \bra{010100} - 
    \bra{101011} -
    \bra{010111} -
    \bra{101000} \nonumber \\  &&~~~~~~~~~~~~~~~~~~~~ -
    \bra{011011} -
    \bra{100100} -
    \bra{011110} -
    \bra{100001}] \Big\}
    \label{eq:phase_6}
\end{eqnarray}
\end{widetext}

\subsection{Bit-phase-flip noisy environment}
The Kraus operators of bit-phase-flip are given by the following matrices given in Eq. \eqref{eqn_bitphase_noise} \cite{fortes2015fighting, oh2002fidelity},

\begin{eqnarray}
 && E_0^F= \sqrt{1-\eta_F} \mathbb{I} =
  \sqrt{1-\eta_F} \begin{pmatrix}
    1 & 0 \\
    0 & 1
  \end{pmatrix}~, \nonumber \\
  && E_1^F= \sqrt{\eta_F}\mathbb{Y} = \sqrt{\eta_F}
  \begin{pmatrix}
    0 & -i \\
    i & 0
  \end{pmatrix}
\label{eqn_bitphase_noise}
\end{eqnarray}
where $0\leq \eta_{F} \leq 1$ represents the bit-phase-flip error probability, which describes the possibility of occurring error in quantum state due to travel qubit. The effect of bit-phase flip noise on the the six entangled channels can be seen by evaluating the affected density matrix after noise has been introduced in the channel, the affected density matrix under the bit-phase flip noise for all the six channels the Bell channel, GHZ channel, two-qubit cluster state, three-qubit cluster state, Brown \emph{et al.} state and the Borras \emph{et al.} are denoted by $\mathcal{E}_1^F (\rho), \mathcal{E}_2^F (\rho), \mathcal{E}_3^F (\rho), \mathcal{E}_4^F (\rho), \mathcal{E}_5^F (\rho), \mathcal{E}_6^F (\rho)$ respectively,  given in Eq. \eqref{eq:bitphase_1} to \eqref{eq:bitphase_6}.

\begin{widetext}
The affected density matrices are influenced by the bit-phase flip noise are given by the following equations from Eq. \ref{eq:bitphase_1} to Eq. \ref{eq:bitphase_6}. Here $\ket{\Psi_6}$ is the Borras \emph{et al.} state given in Eq. \eqref{Borras_Eq1}.
\begin{eqnarray}
&& \mathcal{E}_1^F (\rho) = \frac{1}{2} \Big\{ (1-\eta_F)^2[\ket{00} + \ket{11}][\bra{00} + \bra{11} ] + (\eta_F)^2[\ket{00} - \ket{11} ][\bra{00} - \bra{11} ] \Big\} \label{eq:bitphase_1} \\
&& \mathcal{E}_2^F (\rho) = \frac{1}{2} \Big\{ (1-\eta_F)^3[\ket{000} + \ket{111}][\bra{000} + \bra{111} ] - \eta_F^3[\ket{111} - \ket{000}] [\bra{111} - \bra{000} ] \Big\} \label{eq:bitphase_2} \\
&& \mathcal{E}_3^F (\rho) = \frac{1}{4} \Big\{ [(1-\eta_F)^{2} ][\ket{00} +\ket{01} +\ket{10} - \ket{11}][\bra{00} + \bra{01} + \bra{10} - \bra{11} ] + (\eta_F)^2 [\ket{11} - \ket{01} - \ket{10} - \ket{00}]\nonumber \\ 
&& ~~~~~~~~~~~~~~~~~~~~ [\bra{11} - \bra{01} - \bra{10} - \bra{00} ] \Big\} \label{eq:bitphase_3} \\
&& \mathcal{E}_4^F (\rho) = \frac{1}{8} \Big\{ [(1-\eta_F)^{3} ][\ket{000} +\ket{001} +\ket{010} - \ket{011} + \ket{100} +\ket{101} - \ket{110} + \ket{111}][\bra{000} + \bra{001} + \bra{010}\nonumber \\ 
&&~~~~~~~~~~~~~~~~~~~~ - \bra{011} + \bra{100} + \bra{101} - \bra{110} + \bra{111}] + (i\eta_F)^3 [\ket{111} -\ket{110} -\ket{101} - \ket{100} - \ket{011} +\ket{010} - \ket{001} \nonumber \\ 
&&~~~~~~~~~~~~~~~~~~~~  - \ket{000}][\bra{111} - \bra{110} - \bra{101} - \bra{100} - \bra{011} + \bra{010} - \bra{001} - \bra{000}] \Big\} \label{eq:bitphase_4} \\
&&\mathcal{E}_5^F (\rho) = \frac{1}{8} \Big\{[(1-\eta_F)^{5} ][-\ket{00101} + \ket{00111} + \ket{01000} - \ket{01010} + \ket{10001} + \ket{10011} + \ket{11100} + \ket{11110}] \nonumber \\  
&&~~~~~~~~~~~~~~~~~ [-\bra{00101} + \bra{00111} + \bra{01000} - \bra{01010} + \bra{10001} + \bra{10011} + \bra{11100} + \bra{11110}]  + i(\eta_F)^5 [-\ket{11010}  \nonumber \\ 
&&~~~~~~~~~~~~~~~~~ -  \ket{11000} - \ket{101111} - \ket{10101} + \ket{01110} - \ket{01100} - \ket{00011} + \ket{00001}] [-\bra{11010}  - \bra{11000} - \bra{10111} \nonumber \\ 
&&~~~~~~~~~~~~~~~~~ - \bra{10101} + \bra{01110} - \bra{01100} - \bra{00011} + \bra{00001}]\Big\} \label{eq:bitphase_5} \\
&& \mathcal{E}_6^F (\rho) = \frac{1}{32} \Big\{ (1-\eta_F)^{6} [\ket{\Psi_6}][\bra{\Psi_6}]
    + (\eta_F)^6 
    [\ket{111111} + 
    \ket{000000} +
    \ket{111100} +
    \ket{000011} +
    \ket{111010} +
    \ket{000101} \nonumber \\  &&~~~~~~~~~~~~~~~~~~~~ + 
    \ket{111001} +
    \ket{000110} + 
    \ket{110110} + 
    \ket{001001} +
    \ket{110000} +
    \ket{001111} +
    \ket{101110} +
    \ket{010001} +
    \ket{101101} \nonumber \\  &&~~~~~~~~~~~~~~~~~~~~ + 
    \ket{010010} +
    \ket{100111} + 
    \ket{011000} +
    \ket{100010} +
    \ket{011101} -
    \ket{110101} -
    \ket{001010} -
    \ket{110011} -
    \ket{001100} \nonumber \\  &&~~~~~~~~~~~~~~~~~~~~ -
    \ket{101011} - 
    \ket{010100} -
    \ket{101000} -
    \ket{010111} -
    \ket{100100} -
    \ket{011011} -
    \ket{100001} -
    \ket{011110}]
    [\bra{111111} \nonumber \\  &&~~~~~~~~~~~~~~~~~~~~ + 
    \bra{000000} +
    \bra{111100} +
    \bra{000011} +
    \bra{111010} +
    \bra{000101} +
    \bra{111001} +
    \bra{000110} +
    \bra{110110} +
    \bra{001001} \nonumber \\  &&~~~~~~~~~~~~~~~~~~~~ +
    \bra{110000} +
    \bra{001111} +
    \bra{101110} +
    \bra{010001} +
    \bra{101101} +
    \bra{010010} +
    \bra{100111} +
    \bra{011000} +
    \bra{100010} \nonumber \\  &&~~~~~~~~~~~~~~~~~~~~ + 
    \bra{011101} -
    \bra{110101} -
    \bra{001010} -
    \bra{110011} -
    \bra{001100} -
    \bra{101011} -
    \bra{010100} -
    \bra{101000} -
    \bra{010111} \nonumber \\  &&~~~~~~~~~~~~~~~~~~~~ -
    \bra{100100} - 
    \bra{011011} -
    \bra{100001} -
    \bra{011110}] \Big\}
    \label{eq:bitphase_6}
\end{eqnarray}
\end{widetext}

\subsection{Effect of amplitude damping (AD) noise}
The process of amplitude damping is an essential concept in modelling the energy dissipation in several quantum systems, and the following matrices give its Kraus operators \cite{fortes2015fighting, oh2002fidelity}.
\begin{eqnarray}
  E_0^A=
  \begin{pmatrix}
    1 & 0 \\
    0 & \sqrt{1-\eta_{A}}
  \end{pmatrix}~, \quad
  E_1^A=
  \begin{pmatrix}
    0 & \sqrt{\eta_{A}} \\
    0 & 0
  \end{pmatrix}
\label{eqn_AD_noise}
\end{eqnarray}
Where $0\leq \eta_{A} \leq 1$ represents the decoherence rate of amplitude damping, which describes the possibility of occurring error in the quantum state due to travel qubit. The effect of amplitude damping on the six entangled channels can be seen by evaluating the affected density matrix after the noise has been introduced in the channel, the affected density matrix under the amplitude damping noise for all the six entangled channels- the Bell channel, GHZ channel, two-qubit cluster state, three-qubit cluster state, Brown \emph{et al.} state and the Borras \emph{et al.} are denoted by $\mathcal{E}_1^A (\rho), \mathcal{E}_2^A (\rho), \mathcal{E}_3^A (\rho), \mathcal{E}_4^A (\rho), \mathcal{E}_5^A (\rho), \mathcal{E}_6^A (\rho)$ respectively, given in Eq. \eqref{eq:AD_1} to \eqref{eq:AD_6}.


\begin{widetext}
The affected density matrices are influenced by the amplitude damping noise are given by the following equations from Eq. \eqref{eq:AD_1} to Eq. \eqref{eq:AD_6}
\begin{eqnarray}
    &&\mathcal{E}^A (\rho) = \frac{1}{2} \Big\{ [\ket{00} + (1-\eta_A)\ket{11}] [\bra{00} + (1-\eta_A)\bra{11} ] + (\eta_A)^2[\ket{00}\bra{00}] \Big\} \label{eq:AD_1} \\
&&\mathcal{E}_2^A (\rho) = \frac{1}{2} \Big\{ [\ket{000} + (1-\eta_A)^{\frac{3}{2}}\ket{111}] [\bra{000} + (1-\eta_A)^{\frac{3}{2}}\bra{111} ] + (\eta_A)^3[\ket{000}\bra{000}] \Big\} \label{eq:AD_2} \\
&& \mathcal{E}_3^A (\rho) = \frac{1}{4} \Big\{ [ \ket{00} + \sqrt{1-\eta_A}(\ket{01} + \ket{10} - \ket{11})] [\bra{00} +\sqrt{1-\eta_A}(\bra{01} + \bra{10} - \bra{11})] + (\eta_A)^2[\ket{00}][\bra{00}] \Big\} \label{eq:AD_3} \\
&& \mathcal{E}_4^A (\rho) = \frac{1}{8} \Big\{ 
    [\ket{000} + 
    \sqrt{(1-\eta_A)}(\ket{001} + 
    \ket{010} -
    \ket{100}) + 
    (1-\eta_A)^2(\ket{011} +
    \ket{101} -
    \ket{110}) +
    \sqrt{(1-\eta_A)^3}
    \ket{111}] \nonumber \\  && ~~~~~~~~~~~~~~~~~~~~
    [\bra{000} + 
    \sqrt{(1-\eta_A)}(\bra{001} + 
    \bra{010} -
    \bra{100}) + 
    (1-\eta_A)^2(\bra{011} + 
    \bra{101} -\bra{110}) \nonumber \\  && ~~~~~~~~~~~~~~~~~~~~ +
    \sqrt{(1-\eta_A)^3}
    \bra{111}] + (\eta_A)^3 [\ket{000}][\bra{000}] \Big\} \label{eq:AD_4} \\
&& \mathcal{E}_5^A (\rho) = \frac{1}{8} \Big\{
    [\sqrt{1-\eta_A}\ket{01000} + 
    (1-\eta_A)(-\ket{00101} + \ket{10001} - \ket{01010}) + 
    \sqrt{(1-\eta_A)^3}(\ket{00111} +  \ket{10011} \nonumber \\  && ~~~~~~~~~~~~~~~~~~~~+ 
    \ket{11100}) +
    (1-\eta_A)^2\ket{11110}][\sqrt{1-\eta_A}\bra{01000} + 
    (1-\eta_A)(-\bra{00101} + \bra{10001} - \bra{01010})\nonumber \\  &&~~~~~~~~~~~~~~~~~~~~+ 
    \sqrt{(1-\eta_A)^3}(\bra{00111} +  \bra{10011} + \bra{11100}) +
    (1-\eta_A)^2\bra{11110}]\Big\} \label{eq:AD_5} \\
&& \mathcal{E}_6^A(\rho) = \frac{1}{32} \Big\{ 
    [\ket{000000} + 
    (1-\eta_A)^3\ket{111111} + 
    (1-\eta_A)(\ket{000011} + 
    \ket{000101} + 
    \ket{000110} +
    \ket{001001} + 
    \ket{010001} \nonumber \\ && ~~~~~~~~~~~~~~~~~~~~ +
    \ket{110000} +
    \ket{010010} + 
    \ket{011000} +
    \ket{100010} -
    \ket{001010} -
    \ket{001100} -
    \ket{010100} -
    \ket{101000} -
    \ket{100100} \nonumber \\ &&~~~~~~~~~~~~~~~~~~~~ -
    \ket{100001} ) + (1-\eta_A)^2
    (\ket{111100} + 
    \ket{111010} + 
    \ket{111001} +
    \ket{110110} +
    \ket{001111} + 
    \ket{101110} + 
    \ket{101101} \nonumber \\ &&~~~~~~~~~~~~~~~~~~~~ + 
    \ket{100111} + 
    \ket{011101} +
    \ket{110101} -
    \ket{110011} -
    \ket{101011} +
    +\ket{010111} -
    \ket{011011} - 
    \ket{011110})] 
    [\bra{000000} \nonumber \\ && ~~~~~~~~~~~~~~~~~~~~+ 
    (1-\eta_A)^3\bra{111111} + 
    (1-\eta_A)(\bra{000011} + 
    \bra{000101} + 
    \bra{000110} +
    \bra{001001} +
    \bra{010001} +
    \bra{110000} \nonumber \\ &&~~~~~~~~~~~~~~~~~~~~ +
    \bra{010010} + 
    \bra{011000} +
    \bra{100010} -
    \bra{001010} -
    \bra{001100} -
    \bra{010100} -
    \bra{101000} -
    \bra{100100} -
    \bra{100001}) \nonumber \\ && ~~~~~~~~~~~~~~~~~~~~ + (1-\eta_A)^2
    (\bra{111100} + 
    \bra{111010} + 
    \bra{111001} +
    \bra{110110} +
    \bra{001111} + 
    \bra{101110} + 
    \bra{101101} +
    \bra{100111} \nonumber \\ && ~~~~~~~~~~~~~~~~~~~~ +
    \bra{011101} +
    \bra{110101} -
    \bra{110011} -
    \bra{101011} + 
    \bra{010111} - 
    \bra{011011} - 
    \bra{011110})] \nonumber \\ && ~~~~~~~~~~~~~~~~~~~~ + 
    (\eta_A)^6 [\ket{000000}] [\bra{000000}] \Big\}
    \label{eq:AD_6}
\end{eqnarray}
\end{widetext}

\subsection{Phase damping noisy environment}
Phase damping involves the loss of information about relative phases in a quantum state. During phase damping, the principal quantum system becomes entangled with the environment \cite{DMW2007}. The Kraus operators \{$E_{0}^P,E_{1}^P,E_{2}^P$\} for phase damping noise can be described in Eq. \eqref{eqn_PD_noise} \cite{fortes2015fighting, oh2002fidelity}.
\begin{eqnarray}
  && E_0^P= \sqrt{1-\eta_{P}}
  \begin{pmatrix}
    1 & 0 \\
    0 & 1
  \end{pmatrix}~, \quad
  E_1^P=
  \begin{pmatrix}
    \sqrt{\eta_{P}} & 0 \\
    0 & 0
  \end{pmatrix}~, \quad \nonumber \\
  && E_2^P=
  \begin{pmatrix}
      0 & 0 \\
      0 & \sqrt{\eta_{P}}
  \end{pmatrix}
\label{eqn_PD_noise}
\end{eqnarray}
where $0\leq \eta_{P} \leq 1$ represents the decoherence rate of phase damping, which describes the possibility of occurring error in quantum state due to travel qubit. The effect of phase damping on the the six entangled channels can be seen by evaluating the affected density matrix after noise has been introduced in the channel, the affected density matrix under the phase damping noise for all the six entangled channels- the Bell channel, GHZ channel, two-qubit cluster state, three-qubit cluster state, Brown \emph{et al.} state and the Borras \emph{et al.} are denoted by $\mathcal{E}_1^P (\rho), \mathcal{E}_2^P (\rho), \mathcal{E}_3^P (\rho), \mathcal{E}_4^P (\rho), \mathcal{E}_5^P (\rho), \mathcal{E}_6^P (\rho)$ respectively. They are  given in equation \eqref{eq:PD_1} to \eqref{eq:PD_6}.

\begin{widetext}
The affected density matrices are influenced by the phase damping noise are given by the following equations from Eq. \eqref{eq:PD_1} to Eq. \eqref{eq:PD_6}
\begin{eqnarray}
&& \mathcal{E}_1^P (\rho) = \frac{1}{2} \Big\{ (1-\eta_P)^2 [\ket{00} + \ket{11}] [\bra{00} + \bra{11} ] + (\eta_P)^2[\ket{00}\bra{00} + \ket{11}\bra{11}] \Big\} \label{eq:PD_1} \\
&& \mathcal{E}_2^P (\rho) = \frac{1}{2} \Big\{ (1-\eta_P)^3 [\ket{000} + \ket{111}] [\bra{000} + \bra{111} ] + (\eta_P)^3[\ket{000}\bra{000} + \ket{111}\bra{111}] \Big\} \label{eq:PD_2} \\
&& \mathcal{E}_3^P (\rho) = \frac{1}{4} \Big\{(1-\eta_P)^2[\ket{00} + \ket{01} + \ket{10} - \ket{11}] [\bra{00} + \bra{01} + \bra{10}-\bra{11}] + (\eta_P)^4 [\ket{00}\bra{00}][\ket{11}\bra{11}]  \Big\}  \label{eq:PD_3} \\
&& \mathcal{E}_4^P (\rho) = \frac{1}{8} \Big\{ [(1-\eta_P)^{3}][\ket{000} +\ket{001} +\ket{010} - \ket{011} + \ket{100} + \ket{101} - \ket{110} + \ket{111}] [\bra{000} + \bra{001} + \bra{010} \nonumber \\ 
    &&~~~~~~~~~~~~~~~~~~~~ - \bra{011} + \bra{100} + \bra{101} - \bra{110} + \bra{111}] + (\eta_P)^3 [\ket{000}\bra{000} + \ket{111}\bra{111}] \Big\} \label{eq:PD_4} \\
&& \mathcal{E}_5^P (\rho) = \frac{1}{8} \Big\{(1-\eta_A)^5[-\ket{00101} + \ket{00111} + \ket{01000} - \ket{01010} + \ket{10001} + \ket{10011} + \ket{11100} + \ket{11110}] \nonumber \\  
&&~~~~~~~~~~~~~~~~~~~~ [-\bra{00101} + \bra{00111} + \bra{01000} - \bra{01010} + \bra{10001} + \bra{10011} + \bra{11100} + \bra{11110}]\Big\} \label{eq:PD_5} \\
&& \mathcal{E}_6^P(\rho) = \frac{1}{32} \Big\{ 
    (1-\eta_P)^6[\ket{000000} + 
    \ket{111111} + 
    \ket{000011} + 
    \ket{000101} + 
    \ket{000110} +
    \ket{001001} +
    \ket{010001} + 
    \ket{110000} \nonumber \\  &&~~~~~~~~~~~~~~~~~~~~ +
    \ket{010010} + 
    \ket{011000} +
    \ket{100010} -
    \ket{001010} -
    \ket{001100} -
    \ket{010100} -
    \ket{101000} -
    \ket{100100} -
    \ket{100001} \nonumber \\  &&~~~~~~~~~~~~~~~~~~~~ +
    (\ket{111100} + 
    \ket{111010} + 
    \ket{111001} +
    \ket{110110} +
    \ket{001111} + 
    \ket{101110} + 
    \ket{101101} + 
    \ket{100111} + 
    \ket{011101} \nonumber \\  && ~~~~~~~~~~~~~~~~~~~~ +
    \ket{110101} -
    \ket{110011} -
    \ket{101011} +
    +\ket{010111} -
    \ket{011011} - 
    \ket{011110}]  
    [\bra{000000} + 
    \bra{111111} + 
    \bra{000011}\nonumber \\  &&~~~~~~~~~~~~~~~~~~~~ +
    \bra{000101} + 
    \bra{000110} +
    \bra{001001} +
    \bra{010001} +
    \bra{110000} +
    \bra{010010} + 
    \bra{011000} +
    \bra{100010} -
    \bra{001010}  \nonumber \\  && ~~~~~~~~~~~~~~~~~~~~ -
    \bra{001100} -
    \bra{010100} -
    \bra{101000} -
    \bra{100100} 
   \bra{100001} +
    \bra{111100} + 
    \bra{111010} + 
    \bra{111001} +
    \bra{110110} \nonumber \\  && ~~~~~~~~~~~~~~~~~~~~ + 
    \bra{001111} + 
    \bra{101110} + 
    \bra{101101}  +
    \bra{100111} +
    \bra{011101} +
    \bra{110101} -
    \bra{110011} -
    \bra{101011} + 
    \bra{010111} \nonumber \\  && ~~~~~~~~~~~~~~~~~~~~  -  
    \bra{011011} - 
    \bra{011110}] +
    (\eta_P)^6 [\ket{000000} \bra{000000} + \ket{111111} \bra{111111}] \Big\}
    \label{eq:PD_6}
\end{eqnarray}
\end{widetext}

\subsection{Depolarizing Noisy Environment}
In depolarizing noisy environment, the qubits of the quantum state are depolarized with probability $\eta_D$ and the qubits are left invariant probability $1-\eta_D$. The Pauli operators $\mathbb{X}, \mathbb{Y}$ and $\mathbb{Z}$ act on the qubits with probability $\eta_D/3$. The following matrices give the Kraus operators \cite{fortes2015fighting, oh2002fidelity}.

\begin{eqnarray}
 && E_0^D= \sqrt{1-\eta_{D}}
  \begin{pmatrix}
    1 & 0 \\
    0 & 1
  \end{pmatrix}~, \quad
  E_1^D= \sqrt{\frac{\eta_{D}}{3}}
  \begin{pmatrix}
    0 & 1 \\
    1 & 0
  \end{pmatrix}~, \quad \nonumber \\
  && E_2^D= \sqrt{\frac{\eta_{D}}{3}}
  \begin{pmatrix}
      0 & -i \\
      i & 0
  \end{pmatrix}~, \quad
  E_3^D= \sqrt{\frac{\eta_{D}}{3}}
  \begin{pmatrix}
      1 & 0 \\
      0 & -1
  \end{pmatrix}
\label{eqn_DN_noise}
\end{eqnarray}

The effect of depolarizing noise on the six entangled channels can be seen by evaluating the affected density matrix after noise has been introduced in the channel, the affected density matrix under the depolarizing noise for all the six entangled channels- the Bell channel, GHZ channel, two-qubit cluster state, three-qubit cluster state, Brown \emph{et al.} state and the Borras \emph{et al.} are denoted by $\mathcal{E}_1^D (\rho), \mathcal{E}_2^D (\rho), \mathcal{E}_3^D (\rho), \mathcal{E}_4^D (\rho), \mathcal{E}_5^D (\rho), \mathcal{E}_6^D (\rho)$ respectively, given in equation \eqref{eq:DN_1} to \eqref{eq:DN_6}.

\begin{widetext}
The affected density matrices are influenced by the depolarizing noise are given by the following equations from Eq. \eqref{eq:DN_1} to Eq. \eqref{eq:DN_6}. Here $\ket{\Psi_3},\ket{\Psi_5}$ and $\ket{\Psi_6}$ are the three-qubit cluster state, Brown \emph{et al.} and Borras \emph{et al.} state given in Eq. \eqref{eq:3qCeq}, \eqref{Borras_Eq1} and \eqref{eq:Brown} respectively .
\begin{eqnarray}
&&    \mathcal{E}_1^D (\rho) = \frac{1}{2} \Big\{ [(1-\eta_D)^2 + \frac{\eta_D^2}{3}] [\ket{00} + \ket{11}] [\bra{00} + \bra{11} ] \Big\} \label{eq:DN_1} \\
&& \mathcal{E}_2^D (\rho) = \frac{1}{2} \Big\{ [(1-\eta_D)^3 + \frac{\eta_D}{27}^3 + i\frac{\eta_D}{27}^3] [\ket{000} + \ket{111}][\bra{000} + \bra{111} ] + (\frac{\eta_D^3}{27}) [\ket{000} - \ket{111}][\bra{000} - \bra{111} ] \Big\} \label{eq:DN_2} \\
&& \mathcal{E}_3^D (\rho) = \frac{1}{4} \Big\{ (1-\eta_D)^2 [\ket{00} + \ket{01} + \ket{10} - \ket{11}] [\bra{00} +\bra{01} + \bra{10} - \bra{11}] + 2\frac{\eta_D^2}{9}[\ket{11} + \ket{10} + \ket{01} \nonumber \\  
&& ~~~~~~~~~~~~~~~~~~~~ - \ket{00}][\bra{11} + \bra{10} + \bra{01} - \bra{00}] + \frac{\eta_D^2}{9}[\ket{00} - \ket{01} - \ket{10}-\ket{11}][\bra{00} - \bra{01} - \bra{10} - \bra{11}] \Big\} \label{eq:DN_3} \\
&& \mathcal{E}_4^D (\rho) = \frac{1}{8} \Big\{ [(1-\eta_D)^{3} ][\ket{\Psi_4}][\bra{\Psi_4}] + \Big( \dfrac{\eta_D}{3}     \Big)^3 [\ket{111} +\ket{110} +\ket{101} - \ket{100} + \ket{011} +\ket{010} - \ket{001} + \ket{000}][\bra{111}         \nonumber \\  
&& ~~~~~~~~~~~~~~~~~~~~ + \bra{110} + \bra{101} - \bra{100} + \bra{011} + \bra{010} - \bra{001} + \bra{000}] - \Big(             \dfrac{\eta_D}{3} \Big)^3 [\ket{111} -\ket{110} -\ket{101} - \ket{100} - \ket{011} \nonumber \\ 
&& ~~~~~~~~~~~~~~~~~~~~ + \ket{010} - \ket{001} - \ket{000}][\bra{111} - \bra{110} - \bra{101} - \bra{100} - \bra{011} + \bra{010} - \bra{001} - \bra{000}]  +\Big( \dfrac{\eta_D}{3} \Big)^3[\ket{000} \nonumber \\  
&& ~~~~~~~~~~~~~~~~~~~~ -\ket{001} -\ket{010} - \ket{011} - \ket{100} +\ket{101} - \ket{110} - \ket{111}][\bra{000} - \bra{001} - \bra{010} - \bra{011} - \bra{100} \nonumber\\
&& ~~~~~~~~~~~~~~~~~~~~ + \bra{101} - \bra{110} - \bra{111} \Big\}   \label{eq:DN_4} \\
&& \mathcal{E}_5^D (\rho) = \frac{1}{8} \Big\{[(1-\eta_D)^{5} ][\ket{\Psi_5}][\bra{\Psi_5}] + \Big( \dfrac{\eta_D}{3} \Big)^5 [-\ket{11010} + \ket{11000} + \ket{101111} - \ket{10101} + \ket{01110} + \ket{01100} \nonumber\\  
&& ~~~~~~~~~~~~~~~~~~~~ + \ket{00011} + \ket{00001}] [-\bra{11010} + \bra{11000} + \bra{101111} - \bra{10101} + \bra{01110} + \bra{01100} + \bra{00011} + \bra{00001}] \nonumber\\
&& ~~~~~~~~~~~~~~~~~~~~ + \Big(\dfrac{\eta_D}{3} \Big)^5 [-\ket{00101} - \ket{00111} - \ket{01000} - \ket{01010} + \ket{10001} - \ket{10011} - \ket{11100} + \ket{11110}]  [-\bra{00101} \nonumber\\  
&& ~~~~~~~~~~~~~~~~~~~~ - \bra{00111} - \bra{01000} - \bra{01010} + \bra{10001} - \bra{10011} - \bra{11100} + \bra{11110}] - \Big( \dfrac{\eta_D}{3} \Big)^5 [-\ket{11010}  - \ket{11000} \nonumber\\ 
&& ~~~~~~~~~~~~~~~~~~~~ -  \ket{101111} - \ket{10101} + \ket{01110} - \ket{01100} - \ket{00011} + \ket{00001}] [-\bra{11010}  - \bra{11000} - \bra{10111} \nonumber\\  
&& ~~~~~~~~~~~~~~~~~~~~ - \bra{10101} + \bra{01110} - \bra{01100} - \bra{00011} + \bra{00001}] \Big\} \label{eq:DN_5} \nonumber \\
&& \mathcal{E}_6^D(\rho) = \frac{1}{32} \Big\{ [\ket{\Psi_6}][\bra{\Psi_6}] +
    + \Big(\dfrac{\eta_D}{3} \Big)^6  
    [\ket{111111} + 
    \ket{000000} + 
    \ket{111100} + 
    \ket{000011} + 
    \ket{111010} + 
    \ket{000101} +
    \ket{111001} \nonumber \\  && ~~~~~~~~~~~~~~~~~~~~ +
    \ket{000110} + 
    \ket{110110} + 
    \ket{001001} +
    \ket{110000} +
    \ket{001111} +
    \ket{101110} +
    \ket{010001} +
    \ket{101010} + 
    \ket{010010}   \nonumber \\  && ~~~~~~~~~~~~~~~~~~~~ + 
    \ket{100111} + 
    \ket{011000} +
    \ket{100010} +
    \ket{011101} -
    \ket{110101} -
    \ket{001010} -
    \ket{110011} - 
    \ket{001100} -
    \ket{101011}  \nonumber \\  && ~~~~~~~~~~~~~~~~~~~~ - 
    \ket{010100} - 
    \ket{101000} -
    \ket{010111} - 
    \ket{100100} -
    \ket{011011} -
    \ket{100001} -
    \ket{011110}]
    [\bra{111111} +
    \bra{000000} \nonumber \\  &&  ~~~~~~~~~~~~~~~~~~~~ +
    \bra{111100} +
    \bra{000011} + 
    \bra{111010} +
    \bra{000101} +
    \bra{111001} + 
    \bra{000110} +
    \bra{110110} +
    \bra{001001} +
    \bra{110000} \nonumber \\  && ~~~~~~~~~~~~~~~~~~~~ + 
    \bra{001111} + 
    \bra{101110} +
    \bra{010001} +
    \bra{101010} +
    \bra{010010} +
    \bra{100111} +
    \bra{011000} +
    \bra{100010} +
    \bra{011101} \nonumber \\  && ~~~~~~~~~~~~~~~~~~~~ -
    \bra{1100101} - 
    \bra{001010} - 
    \bra{110011} -
    \bra{001100} -
    \bra{101011} -
    \bra{010100} -
    \bra{101000} -
    \bra{010111} -
    \bra{100100} \nonumber \\  && ~~~~~~~~~~~~~~~~~~~~ -
    \bra{011011} -
    \bra{100001} -
    \bra{011110}] + 
    \Big(\dfrac{\eta_D}{3} \Big)^6 
    [\ket{111111} + 
    \ket{000000} + 
    \ket{111100} + 
    \ket{000011} + 
    \ket{111010} \nonumber \\  && ~~~~~~~~~~~~~~~~~~~~ +  
    \ket{000101} + 
    \ket{111001} +
    \ket{000110} + 
    \ket{110110} + 
    \ket{001001} + 
    \ket{110000} + 
    \ket{001111} +
    \ket{101110} +
    \ket{010001} \nonumber\\  && ~~~~~~~~~~~~~~~~~~~~ + 
    \ket{101101} +
    \ket{010010} +
    \ket{100111} + 
    \ket{011000} + 
    \ket{100010} + 
    \ket{011101} - 
    \ket{110101} - 
    \ket{001010} - 
    \ket{110011} \nonumber \\  && ~~~~~~~~~~~~~~~~~~~~ -  
    \ket{001100} - 
    \ket{101011} -
    \ket{010100} - 
    \ket{101000} - 
    \ket{010111} - 
    \ket{100100} - 
    \ket{011011} - 
    \ket{100001} - 
    \ket{011110}] \nonumber \\  && ~~~~~~~~~~~~~~~~~~~~  
    [\bra{111111} +
    \bra{000000} +
    \bra{111100} + 
    \bra{000011} + 
    \bra{111010} + 
    \bra{000101} + 
    \bra{111001} +
    \bra{000110} +
    \bra{110110} \nonumber \\  && ~~~~~~~~~~~~~~~~~~~~ + 
    \bra{001001} +
    \bra{110000} +
    \bra{001111} +
    \bra{101110} +
    \bra{010001} +
    \bra{101101} +
    \bra{010010} +
    \bra{100111} +
    \bra{011000} \nonumber \\  && ~~~~~~~~~~~~~~~~~~~~ + 
    \bra{100010} +
    \bra{011101} -
    \bra{110101} - 
    \bra{001010} -
    \bra{110011} -
    \bra{001100} -
    \bra{101011} -
    \bra{010100} -
    \bra{101000} \nonumber \\   && ~~~~~~~~~~~~~~~~~~~~ -
    \bra{010111} -
    \bra{100100} -
    \bra{011011} -
    \bra{100001} -
    \bra{011110}]
    + \Big(\dfrac{\eta_D}{3} \Big)^6 
    [\ket{000000} +
    \ket{111111} +
    \ket{000011} \nonumber \\  && ~~~~~~~~~~~~~~~~~~~~ + 
    \ket{111100} +
    \ket{000101} + 
    \ket{111010} +
    \ket{000110} +
    \ket{111001} + 
    \ket{001001} + 
    \ket{110110} + 
    \ket{001111} + 
    \ket{110000} \nonumber \\   && ~~~~~~~~~~~~~~~~~~~~ + 
    \ket{010001} + 
    \ket{101110} +
    \ket{010010} +
    \ket{101101} + 
    \ket{011000} + 
    \ket{100111} + 
    \ket{011101} + 
    \ket{100010} +
    \ket{010010} \nonumber \\  && ~~~~~~~~~~~~~~~~~~~~ -   
    \ket{110101} -
    \ket{001100} - 
    \ket{110011} -
    \ket{010100} - 
    \ket{101011} - 
    \ket{010111} - 
    \ket{101000} - 
    \ket{011011} - 
    \ket{100100} \nonumber \\  && ~~~~~~~~~~~~~~~~~~~~ -  
    \ket{011110} -
    \ket{100001}]
    [\bra{000000} +
    \bra{111111} +
    \bra{000011} + 
    \bra{111100} + 
    \bra{000101} + 
    \bra{111010} + 
    \bra{000110} \nonumber \\  && ~~~~~~~~~~~~~~~~~~~~ +  
    \bra{111001} +
    \bra{001001} +
    \bra{110110} +
    \bra{001111} +
    \bra{110000} +
    \bra{010001} +
    \bra{101110} +
    \bra{010010} +
    \bra{101101} \nonumber \\ && ~~~~~~~~~~~~~~~~~~~~ + 
    \bra{011000} +
    \bra{100111} +
    \bra{011101} +
    \bra{100010} +
    \bra{001010} - 
    \bra{110101} -
    \bra{001100} -
    \bra{110011} -
    \bra{010100} \nonumber \\  && ~~~~~~~~~~~~~~~~~~~~ -
    \bra{101011} -
    \bra{010111} -
    \bra{101000} -
    \bra{011011} -
    \bra{100100} -
    \bra{011110} -
    \bra{100001}] \Big\}
    \label{eq:DN_6}
\end{eqnarray}
\end{widetext}

\section{Discussion and Conclusion }\label{Sec9}
In this study, the complexity in teleportation is studied among six different entangled channels under the effect of six different kinds of noises. The teleportation is performed via all four Bell channels, all eight GHZ channels, all four two-qubit cluster states channel, all eight three-qubit cluster state channels, the five qubits highly entangled Brown \emph{et al.} channel and the six qubit Borras \emph{et al.} channel. The quantum cost is calculated for all the teleportation protocols and visually represented in Fig. \ref{fig:Q Cost}. 

Then the effect of noise on the entangled channel is studied. We have derived the density matrices of the teleported state, which is influenced by the noise. The fidelity provides the closeness between the two quantum states. The effect of bit-flip noise is visualized with the help of graphical representation of variation in the fidelity against the phase-flip noise parameter $\eta_B$ given in  Fig. \ref{fig:Graph Plot}. From the graph, it is illustrated that fidelity decreases with the increase in the noise parameter $\eta_B \in [0,0.5]$. After that, the fidelity shows an upward trend and reaches one as $\eta_W \in [0.5,1]$ in case Bell channel, GHZ channel, three-qubit cluster sate, Brown \emph{et al.} and Borras \emph{et al.}, but in two-qubit cluster channel, the fidelity decreases rapidly to zero as the noise parameter $\eta_B \in [0.5,1]$. The effect of phase-flip noise is visualized with the help of graphical representation of variation in the fidelity against the bit-flip noise parameter $\eta_W$ is given in  Fig. \ref{fig:Graph Plot}. From the plot, it is illustrated that fidelity decreases with the increase in the noise parameter $\eta_W \in [0,0.5]$. After that, the fidelity shows an upward trend and reaches one as $\eta_W \in [0.5,1]$ in case Bell channel, Brown \emph{et al.} and Borras \emph{et al.} but in other three channels viz. GHZ channel, two-qubit cluster state and three-qubit cluster state the fidelity increases but does not reach one as the noise parameter $\eta_W \in [0.5,1]$. In the case of bit-phase flip noise, it is visualized with the help of graphical representation of variation in the fidelity against the bit-phase flip noise parameter $\eta_F$ is given in  Fig. \ref{fig:Graph Plot}. From the plot, it is illustrated that fidelity decreases with the increase in the noise parameter $\eta_F \in [0,0.5]$. After that, the fidelity shows an upward trend and reaches one as $\eta_F \in [0.5,1]$ in case two-qubit cluster state and Brown \emph{et al.} but in other entangled channels increases slowly but does not reach one as the noise parameter increases in the range $\eta_F \in [0.5,1]$. For amplitude damping noise, the graphical representation of variation in the fidelity against the amplitude damping noise parameter $\eta_A$ is given in  Fig. \ref{fig:Graph Plot}. From the plot, it is illustrated that fidelity decreases with the increase in the noise parameter $\eta_A \in [0,0.5]$, after that the fidelity shows an upward trend as $\eta_A \in [0.5,1]$ in case Bell channel and GHZ channel but in all other entangled channels the fidelity decreases monotonically as the noise parameter increases in the range $\eta_A \in [0.5,1]$, in-fact it reaches zero for $\eta_A=1$ in case of teleportation via Brown \emph{et al.} state, represented in Fig. \ref{fig:Graph Plot}(e). For phase damping noise, the graphical representation of variation in the fidelity against the phase damping noise parameter $\eta_P$ is given in  Fig. \ref{fig:Graph Plot}. From the plot, it is illustrated that fidelity decreases with the increase in the noise parameter $\eta_P \in [0,0.5]$, after that the fidelity shows an upward trend as $\eta_P \in [0.5,1]$ in case Bell channel and GHZ channel, and a slight increase in the case of three-qubit cluster state as can be seen in Fig. \ref{fig:Graph Plot}(d). However, in all other entangled channels, the fidelity decreases monotonically as the noise parameter increases in the range $\eta_P \in [0.5,1]$, in-fact it reaches zero for $\eta_P=1$ in case of teleportation via two-qubit cluster state and Brown \emph{et al.} state. One more interesting observation is that the effect of phase flip noise coincides with the bit-flip and phase-flip noises in case of teleportation via the two-qubit cluster state, as can be seen in Fig. \ref{fig:Graph Plot}(c). For depolarizing noise, the graphical representation of variation in the fidelity against the phase damping noise parameter $\eta_D$ is given in  Fig. \ref{fig:Graph Plot}. From the plot, it is illustrated that fidelity decreases with the increase in the noise parameter $\eta_D \in [0,0.5]$. After that, the fidelity shows a slight upward trend as $\eta_D \in [0.5,1]$ in case of teleportation via Bell channel as can be seen in Fig. \ref{fig:Graph Plot}(a). However, in all other entangled channels, the fidelity decreases monotonically as the noise parameter increases in the range $\eta_D \in [0.5,1]$. In fact it reaches zero for $\eta_D=1$ in case of teleportation via Borras \emph{et al.} state. One more interesting observation is that the effect of depolarizing noise coincides with the phase flip noise in the case of teleportation via the Brown \emph{et al.} state. In a nutshell, the amplitude damping, phase damping and depolarizing noise have the maximum impact on teleportation. 
It is cumbersome to perform its practical implementation on real quantum computers due to non-unitary operators involved in the noise implementation process. However, this is attainable and can be achieved in the near future as a potential application.




\end{document}